\pdfoutput=1

\documentclass[11pt,twoside,a4paper,cmspaper,final,collab]{cms-tdr}
\def\svnVersion{267485}\def\cmsCernNo{2013-037}\def\cmsCernDate{\today}\def\cmsMessage{Published in Physical Review D as \href{http://dx.doi.org/10.1103/PhysRevD.86.052013}{\doi{10.1103/PhysRevD.86.052013.}}}

\begin{document}\cmsNoteHeader{EXO-11-028}

\hyphenation{had-ron-i-za-tion}
\hyphenation{cal-or-i-me-ter}
\hyphenation{de-vices}

\RCS$Revision: 145348 $
\RCS$HeadURL: svn+ssh://svn.cern.ch/reps/tdr2/papers/EXO-11-028/trunk/EXO-11-028.tex $
\RCS$Id: EXO-11-028.tex 145348 2012-08-30 01:15:31Z emanuela $
\newcommand{\ST}{\ensuremath{S_{\mathrm{T}}}\xspace}
\newcommand{\MT}{\ensuremath{M_{\mathrm{T}}}\xspace}
\ifthenelse{\boolean{cms@external}}{\providecommand{\cmsLeft}{top}}{\providecommand{\cmsLeft}{left}}
\ifthenelse{\boolean{cms@external}}{\providecommand{\cmsLLeft}{Top}}{\providecommand{\cmsLLeft}{Left}}
\ifthenelse{\boolean{cms@external}}{\providecommand{\cmsRight}{bottom}}{\providecommand{\cmsRight}{right}}
\cmsNoteHeader{EXO-11-028} % This is over-written in the CMS environment: useful as preprint no. for export versions
\title{Search for pair production of first- and second-generation scalar leptoquarks in \texorpdfstring{\Pp\Pp\ collisions at $\sqrt{s}= 7\TeV$}{pp collisions at sqrt(s) = 7 TeV}}

\date{\today}

\abstract{
Results are presented from a search for the pair production of first- and second-generation scalar
leptoquarks in proton-proton collisions at $\sqrt{s}= 7\TeV$. The data sample corresponds to an integrated luminosity of 5.0\fbinv, collected by the CMS detector at the LHC. The search signatures involve either two charged leptons of the same flavor (electrons or muons) and at least two jets, or a single charged lepton (electron or muon), missing transverse energy, and at least two jets.
If the branching fraction of the leptoquark decay into a charged lepton and a quark is assumed to be $\beta=1$, leptoquark pair production is excluded at the 95\% confidence level for masses below 830\GeV and 840\GeV for the first and second generations, respectively. For $\beta = 0.5$, masses below 640\GeV and 650\GeV are excluded. These limits are the most stringent to date.}

\hypersetup{%
pdfauthor={CMS Collaboration},%
pdftitle={Search for pair production of first- and second-generation scalar leptoquarks in pp collisions at sqrt(s)= 7 TeV},%
pdfsubject={CMS},%
pdfkeywords={CMS, physics, leptoquarks}}

\maketitle %maketitle comes after all the front information has been supplied

\section{Introduction}
\label{intro}
\begin{figure*}[htbp]
       \begin{center}
       {\includegraphics[width=.3\textwidth]{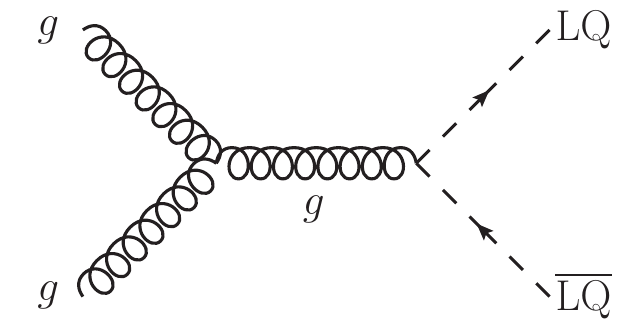}}
       {\includegraphics[width=.3\textwidth]{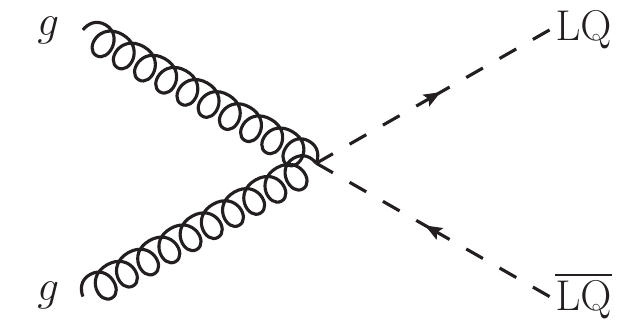}}\\
       {\includegraphics[width=.3\textwidth]{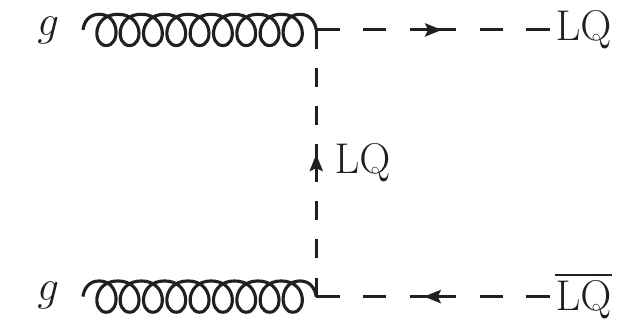}}
       {\includegraphics[width=.3\textwidth]{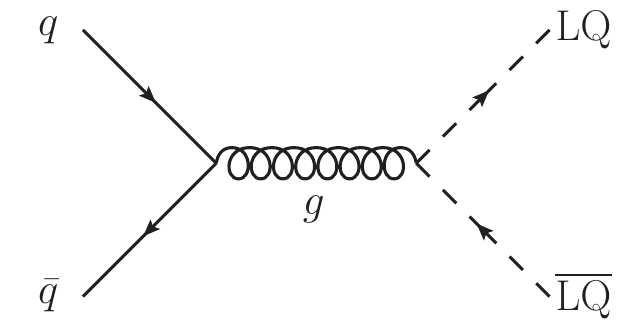}}
       \caption{Dominant leading order diagrams for the pair production of scalar leptoquarks.}
	  \label{fig:LQFeynman}
	  \end{center}
\end{figure*}

The structure of the standard model (SM) of particle physics suggests
a fundamental relationship between quarks and leptons.
There are many models beyond the SM that predict the
existence of leptoquarks (LQ), hypothetical particles that carry both baryon number and lepton number and couple to both quarks and leptons.
Among these scenarios are grand unified theories~\cite{gut1,gut2}, composite models~\cite{Buchmuller:1986zs},
extended technicolor models~\cite{techni1,techni2,techni3},
and superstring-inspired models~\cite{superstring}.
Leptoquarks are color triplets with fractional electric charge, and can be either scalar or vector particles. A leptoquark couples to a lepton and a quark with a coupling strength $\lambda$, and it decays to a charged lepton and a quark with an unknown branching fraction $\beta$ or to a neutrino and a quark with branching fraction {$1-\beta$}. In order to satisfy constraints from bounds on flavor-changing neutral currents and from rare pion and kaon decays~\cite{Buchmuller:1986zs,FCNC}, it is assumed that leptoquarks couple to quarks and leptons of a single generation. Leptoquarks are classified as \mbox{first-,}~\mbox{second-,} or third-generation, depending on the generation of leptons to which they couple.
The dominant mechanisms for the production of leptoquark pairs at the Large Hadron Collider (LHC) are gluon-gluon ($\Glu\Glu$) fusion and quark-antiquark ($\cPq\cPaq$) annihilation, shown in Fig.~\ref{fig:LQFeynman}.
The dominant processes only depend on the strong coupling constant and have been calculated at next-to-leading order (NLO)~\cite{PhysRevD.71.057503}. The cross section for production via the unknown Yukawa coupling
$\lambda$ of a leptoquark to a lepton and a quark is typically smaller.

This paper reports on a search for pair production of scalar leptoquarks. Several experiments have searched for pair-produced scalar leptoquarks but none has obtained evidence for them ~\cite{D0,CMSLQ,CMSLQ1,CMSLQ2,ATLASLQ2,ATLASLQ1}. This search uses a data sample corresponding to an integrated luminosity of 5.0\fbinv recorded with the Compact Muon Solenoid (CMS) detector during the 2011 proton-proton run of the LHC at $\sqrt{s}=7$\TeV. The analysis performed in this paper considers the decay of leptoquark pairs into two charged leptons of the same flavor (either electrons or muons) and two quarks; or into a charged lepton, a neutrino, and two quarks. As a result, two distinct classes of events are selected: one with two high-transverse-momentum ($\PT$) electrons or muons and at least two high-$\PT$ jets ($\ell\ell jj$) and the other with one high-$\PT$ electron or muon, large missing transverse energy ($\MET$), and at least two high-$\PT$ jets ($\ell\nu jj$).

The CMS detector, described in detail elsewhere~\cite{CMS}, uses a cylindrical coordinate system with the $z$ axis along the
counterclockwise beam axis. The detector consists of an inner tracking system and electromagnetic (ECAL) and hadron (HCAL) calorimeters surrounded by a 3.8\unit{T} solenoid. The inner tracking system consists of a silicon pixel and strip tracker, providing the required granularity and precision for the reconstruction of vertices of charged particles in the range $0 \leq \phi \leq 2\pi$ in azimuth and $|\eta|<2.5$, where the pseudorapidity $\eta$ is defined as $\eta =
-\ln[\tan(\theta/2)]$, and $\theta$ is the polar angle measured with respect to the $z$ axis. The crystal ECAL and the brass/scintillator sampling HCAL are used to measure with high resolution the energies of photons, electrons, and hadrons for $|\eta|<3.0$.
The three muon systems surrounding the solenoid cover a region $|\eta|<2.4$ and are composed of drift tubes in the barrel region $(|\eta|<1.2)$, of cathode strip chambers in the endcaps $(0.9<|\eta|<2.4)$, and of resistive plate chambers in both the barrel region and the endcaps $(|\eta|<1.6)$.
Events are recorded based on a trigger decision using information from either the calorimeter or muon systems. The final trigger decision is based on the information from all subsystems, which is passed on to the high-level trigger (HLT), consisting of a farm of computers running a version of the reconstruction software optimized for fast processing.

The $\ell\ell jj$ and $\ell\nu jj$ analyses are performed separately and the results are combined as a function of the branching fraction $\beta$ and the leptoquark mass $M_{\mathrm {LQ}}$ for first and second generations independently. The analysis in all four decay channels searches for leptoquarks in an excess of events characteristic of the decay of heavy objects.
Various triggers are used to collect events depending on the decay channel and the data taking periods as described in Section~\ref{dataobject}.
An initial selection isolates events with high-$\PT$ final-state particles (two or more isolated leptons and two or more jets; or one isolated lepton, two or more jets, and large $\MET$ indicative of the emission of a neutrino). Kinematic variables are then identified to further separate a possible leptoquark signal from the expected backgrounds, and optimized thresholds on the values of these variables are derived in order to maximize the sensitivity to the possible presence of a signal in each decay mode. The variables used in the optimization are the invariant mass of jet-lepton pairs ($M_{\ell j}$), the scalar sum ($\ST$) of the $\PT$ of each of the final state objects, and either the invariant mass of the dilepton pair ($M_{\ell\ell}$) in the $\ell\ell jj$ channels or $\MET$ in the $\ell\nu jj$ channels.

Major sources of SM background are $\cPZ/\gamma^*$+jets, \PW+jets processes, and $\ttbar$. Smaller contributions arise from single top production, diboson processes, and QCD multijet processes. The major backgrounds are determined either from control samples in data or from Monte Carlo (MC) simulated samples normalized to data in selected control regions.

After final selection, the data are well described by the SM background predictions, and upper limits on the leptoquark pair-production cross section are set using a CL$_\mathrm{S}$ modified frequentist approach~\cite{Junk:1999kv,Read:2000ru}. Using Poisson statistics,
 95\% confidence level (CL) upper limits are obtained on the leptoquark pair-production cross section times branching fraction as a function of leptoquark mass ($M_{\mathrm {LQ}}$). This is compared with the NLO predictions~\cite{PhysRevD.71.057503} to determine lower limits on $M_{\mathrm {LQ}}$ for $\beta = 1$ and $\beta = 0.5$. The $\ell\ell jj$ and $\ell\nu jj$ channels are combined to further maximize the exclusion in $\beta$ and $M_{\mathrm {LQ}}$, especially for the case $\beta \sim 0.5$, where combining the two channels increases the sensitivity of the search.

The data and the initial event selection are detailed in Section~\ref{dataobject} of this paper, followed by a description of the signal modeling and background estimates in Section~\ref{model} and Section~\ref{background}, respectively. Section~\ref{selection} contains the final event selection, and Section~\ref{systematics} describes the systematic uncertainties. The results of the search are presented in Section~\ref{results} and summarized in Section~\ref{conclusions}.

\section{Dataset and object reconstruction}
\label{dataobject}
For the first-generation $\Pe\Pe jj$ analysis, events are required to pass a double-electron trigger or a double-photon trigger, with an electron or photon $\PT > 33$\GeV.  For the first-generation $\Pe \nu jj$ analysis, events are required to pass either a single-electron trigger or a trigger based on the requirement of one electron with $\PT$ threshold between 17 and 30\GeV, missing transverse energy threshold between 15 and 20\GeV, and two jets with $\PT$ threshold between 25 and 30\GeV. The trigger thresholds vary according to the run period. For the second-generation leptoquark analyses, events are required to pass a single-muon trigger without isolation requirements and with a $\PT$ threshold of 40\GeV. For the $\Pe\Pe jj$ channel, the trigger efficiency is greater than 99$\%$. For the $\Pe \nu jj$ channel, the electron trigger efficiency is measured to be $95\%$. For the $\mu\mu jj$ and $\mu\nu jj$ channels, the single muon trigger efficiency is measured to be $92\%$ per muon.

Electron candidates~\cite{EMId} are required to have an electromagnetic cluster
with $\PT>$ 40\GeV and pseudorapidity $|\eta|< 2.5$ (2.2)
for the $\Pe\Pe jj$ ($\Pe\nu jj$) analysis, excluding the transition region between the barrel and the endcap
detectors, $1.44<|\eta|<1.57$.
The $\Pe \nu jj$ analysis requires lower electron $|\eta|$ to reduce the QCD multijet background,
with negligible reduction of the signal acceptance.
Electron candidates are required to have an electromagnetic cluster in the ECAL
that is spatially matched to a reconstructed track in the central tracking system in both $\eta$ and the azimuthal angle $\phi$, and
to have a shower shape consistent with that of an electromagnetic shower.
Electron candidates are further required to be isolated from additional energy deposits
in the calorimeter and from reconstructed tracks beyond the matched track in the central tracking system.
In addition, to reject electrons coming from photon conversion in the tracker material,
the track associated with the reconstructed electron is required to have hits in all inner tracker layers.

Muons are reconstructed as tracks in the muon system that are matched to the tracks reconstructed in the inner tracking system~\cite{MuId}. Muons are required to have $\PT > 40$\GeV, and to be reconstructed in the HLT fiducial volume, \ie with $|\eta| < 2.1$. In addition, muons must be isolated by requiring that the tracker-only relative isolation be less than 0.1. Here, the relative isolation is defined as a sum of the transverse momenta
of all tracks in the tracker in a cone of {$\Delta R = \sqrt{(\Delta \phi )^2 + (\Delta \eta) ^2} = 0.3$}
around the muon track (excluding the muon track), divided by the muon $\PT$. To have a precise measurement of the transverse impact parameter of the muon track relative to the beam spot, only muons with tracks containing more than 10 hits in the silicon tracker and at least one hit in the pixel detector are considered. To reject muons from cosmic rays, the transverse impact parameter with respect to the primary vertex is required to be less than 2\unit{mm}.

Jets and $\MET$ are reconstructed using a particle-flow algorithm~\cite{pf_algos}, which identifies and measures stable particles by combining information from all CMS sub-detectors. The $\MET$ calculation uses calorimeter estimates improved by high precision inner tracking information as well as corrections based on particle-level information in the event.  Jets are reconstructed using the anti-$k_\mathrm{T}$~\cite{antikt} algorithm with a
distance parameter of $R = 0.5$. The jet energy is calibrated using $\PT$ balance of dijet and $\gamma+$jet events~\cite{JetCorr}.
In the $\Pe\Pe jj$ and $\mu \mu j j$ ($\Pe\nu jj$ and $\mu \nu j j$) channels, jets are required to have $\PT >30$ (40)\GeV, and $|\eta| < 2.4$.
Furthermore, jets are required to have a spatial separation from electron or muon candidates of $\Delta R > 0.3$.

The initial selection of $\Pe\Pe jj$ or $\mu\mu jj$ events requires two electrons or two muons and at least two jets satisfying the conditions described above. The two leptons
and the two highest-$\PT$ jets are selected as the decay products
from a pair of leptoquarks. The invariant mass of the two electrons (muons) is required to be $M_{\ell\ell}>60~(50)$\GeV. To reduce the combinatorial background, events with a scalar transverse energy
$\ST^{\ell\ell}= \PT(\ell_1) + \PT(\ell_2) + \PT(j_1) + \PT(j_2)$ below 250\GeV are rejected.

For the $\Pe \nu jj~(\mu \nu jj)$ initial selection, events are required to contain one electron (muon) satisfying the conditions described above and at least two jets with $\PT >40$\GeV and $\MET > 55$\GeV. The jet $\PT$ threshold is higher than that in the dilepton channels to account for jet $\PT$ thresholds in triggers used in the $\Pe \nu jj$ channel.
A veto on the presence of extra muons (electrons) is also applied. The angle in the transverse plane between the leading $\PT$ jet and the $\MET$ vector is required to be $\Delta \phi(\MET,j_1) > 0.5$ to reject events with misreconstructed $\MET$. In order to reduce the contribution from QCD multijet events, the lepton and the $\MET$ are required to be separated by $\Delta \phi(\MET,l) > 0.8$. In addition, events are rejected if the scalar transverse energy $\ST^{\ell\nu}= \PT(\ell) + \MET + \PT(j_1) + \PT(j_2)$ is below 250\GeV.

The initial selection criteria are summarized in Table~\ref{tab:EventSel}.

\begin{table} [htb]
     \topcaption{Initial selection criteria in the $\Pe \Pe j j$, $\mu \mu j j$, $\Pe \nu j j$, and $\mu \nu j j$ channels.}
     \label{tab:EventSel}
 \begin{center}
\begin{tabular}{| c  c  c  c  c |} \hline
Variable & $\Pe\Pe j j$& $\mu \mu j j$& $\Pe \nu j j$ &$\mu \nu j j$ \\  \hline\hline
$\PT(\ell_1)$ [\GeVns{}] & $>40$ & $>40$ & $>40$ & $>40$  \\
$\PT(\ell_2)$[\GeVns{}] & $>40$ & $>40$ & --- & ---  \\
$|\eta(\ell_1)|$& $<2.5$ & $<2.1$ & $<2.2$ & $<2.1$  \\
$|\eta(\ell_2)|$ & $<2.5$ & $<2.1$ & --- & ---  \\
$\PT(j_1)$ [\GeVns{}] & $>30$ & $>30$ & $>40$ & $>40$  \\
$\PT(j_2)$ [\GeVns{}] & $>30$ & $>30$ & $>40$ & $>40$  \\
$\Delta R (\ell,j)$ & $>0.3$ & $>0.3$ & $>0.3$ & $>0.3$ \\
$\MET$ [\GeVns{}] & --- & --- & $>55$ & $>55$ \\
$|\Delta \phi(\MET,j_1)| $ & --- & --- & $>0.5$ & $>0.5$ \\
$|\Delta \phi(\MET,\ell)|$ & --- & --- & $>0.8$ & $>0.8$ \\
$M_{\ell\ell}$ [\GeVns{}] & $>$60 & $>50$ & --- & --- \\
$\MT^{\ell\nu}$ [\GeVns{}] & --- & --- & $>50$ & $>50$ \\
$\ST^{\ell\ell}$ [\GeVns{}] & $>250$ & $>250$ & --- & ---  \\
$\ST^{\ell\nu}$ [\GeVns{}]& --- & --- & $>250$ & $>250$ \\\hline
\end{tabular}
\end{center}
 \end{table}

\section{Signal and background modeling}
\label{model}
The MC samples for the signal processes are generated for a range of leptoquark mass hypotheses between {250} and {900\GeV}, with a renormalization and factorization scale $\mu \approx M_{\mathrm {LQ}}$.
The MC generation uses the \PYTHIA generator~\cite{PYTHIA} (version {6.422})
and CTEQ6L1 parton distribution functions (PDF)~\cite{CTEQ}. The MC samples used to estimate the contribution from SM background processes are $\ttbar$+jets events, generated with \MADGRAPH~\cite{MADGRAPH1,MADGRAPH2}; single-top events ($s$, $t$, and $t\PW$ channels), generated with \POWHEG~\cite{POWHEG}; $\cPZ/\gamma^*$+jets events and \PW+jets events, generated with {\SHERPA~\cite{Gleisberg:2008ta}; $VV$ events, where $V$ either represents a \PW\ or a \cPZ\ boson, generated with \PYTHIA; QCD muon-enriched multijet events, generated with \PYTHIA in bins of transverse momentum of the hard-scattering process %$\hat{p}_T$%
from 15\GeV to the kinematic limit.
The simulation of the CMS detector is based on \GEANTfour~\cite{GEANT4}, and includes multiple collisions in a single bunch crossing corresponding to the luminosity profile of the LHC during the data taking periods of interest.

\section{Background estimate}
\label{background}
The main processes that can mimic the signature of
a leptoquark signal in the $\ell\ell jj$ channels are $\cPZ/\gamma^*$+jets,
$\ttbar$, $VV$+jets, \PW+jets,
and QCD multijets.
The $\cPZ/\gamma^*$+jets background is determined by comparing events from data and MC samples
in two different regions: in the region of low (L) dilepton invariant mass around the Z boson mass ($70 < M_{\ell\ell} < 100$\GeV for electrons and $80 < M_{\ell\ell} < 100$\GeV for muons)
and in the region of high (H) mass $M_{\ell\ell} > 100$\GeV. The low mass scaling factor $R_{\cPZ} = {N_{L}}/{N^{\mathrm {MC}}_{L}}$ is measured to be $1.27  \pm  0.02$ for the $\Pe\Pe jj$ channel and $1.29  \pm  0.02$ for the $\mu\mu jj$ channel, where $N_{L}$ and $N^{\mathrm {MC}}_{L}$ are the number of
data and MC events, respectively, in the $\cPZ$ mass window.
The number of $\cPZ/\gamma^*$+jets events above 100\GeV is then
estimated as:
\begin{equation}
N_{H}  = R_{\cPZ} N^{\mathrm {MC}}_{H},
\label{eq:zjets}
\end{equation}
where $N^{\mathrm {MC}}_{H}$ is the number of MC events with $M_{\ell\ell} > 100$\GeV. The estimated number of $\cPZ/\gamma^*$+jets events
is obtained with the selection criteria optimized for different leptoquark mass hypotheses and it is used in the limit setting procedure.

The number and kinematic distributions for the $\ttbar$ events with two leptons of the same flavor is estimated from the number of data events that contain one electron and one muon. This type of background is expected to produce the $\Pe\Pe$ final state or the $\mu\mu$ final state with half the probability of the
$\Pe\mu$ final state. In the data the number of $\Pe\Pe$ or $\mu\mu$ events is estimated to be:
\begin{equation}
N_{\Pe\Pe( \mu \mu )} = \frac{1}{2} \times \frac{\epsilon_{\Pe( \mu )}}{\epsilon_{\mu (\Pe)}} \times  \frac{\epsilon^{\mathrm {trig}}_{\mathrm {ee}( \mu \mu )}}{\epsilon^{\mathrm {trig}}_{\Pe \mu }} \times N_{\Pe \mu },
\label{eqn:1}
\end{equation}
where $\epsilon_{\mu}$ and $\epsilon_{\Pe}$ are the muon and electron reconstruction and identification efficiencies and $\epsilon^{\mathrm {trig}}$ are the HLT efficiencies to select \Pe\Pe, $\mu\mu$, and $\Pe \mu$ events.

No QCD multijet MC events pass the $\mu\mu jj$ final selection. A crosscheck made using a data control sample containing same-sign muons confirms that the QCD multijet background is negligible in this channel.

For the first-generation leptoquark analyses the multijet background contribution is estimated from a data control sample as follows.
The probability that an electron candidate passing
loose electron requirements additionally passes all electron
requirements is measured as a function of $\PT$ and $\eta$ in a data
sample with one and only one electron candidate, two or more jets and
low $\MET$. This sample is dominated by QCD multijet events and
is similar in terms of jet activity to the $\Pe\Pe jj$ and $\Pe \nu jj$ analysis samples.
A correction for a small contamination of genuine electrons passing all electron requirements is derived
from MC simulations.
The QCD multijet background in the final $\Pe\Pe jj$ ($\Pe\nu jj$) selection
is predicted by applying twice (once) the above probability to a
sample with two electron candidates (one electron candidate and large
$\MET$), and two or more jets, which satisfy all the requirements of
the signal selections. The resulting estimate is ${\sim}1\%$ (${\sim}8\%$)
of the total background for the selections corresponding to the region of leptoquark masses
where the exclusion limits are placed.

Contributions to the $\ell\ell jj$ background from $VV$+jets processes and single-top production are small and they are estimated using MC simulation.

In the $\ell\nu jj$ channel, the main backgrounds come from three sources: processes that lead to the production of a genuine \PW\ boson such as \PW+jets,
$\ttbar$, single-top production, diboson processes ($\PW\PW$, $\PW\cPZ$); instrumental background, mostly caused by the misidentification of jets as leptons in multijet
processes, thus creating misidentified electrons or muons and misreconstructed $\MET$ in the final state; and \cPZ\ boson production, such as $\cPZ/\gamma^*$+jets and $\cPZ{}\cPZ{}$ processes.
The contribution from the principal backgrounds, \PW+jets and $\ttbar$, is estimated with MC simulation normalized to data
at the initial selection level in the region $50<\MT<110$\GeV, where $\MT$ is the transverse mass calculated from the lepton $\PT$ and the $\MET$.

The region $50<\MT<110$\GeV is used to determine both the \PW+jets and the $\ttbar$ normalization factors using
two mutually exclusive selections (less than four jets or at least four jets with $\PT>40$\GeV and $|\eta|<2.4$) that separately enhance the samples with \PW+jets and with $\ttbar$ events. The results of these two selections are used to form a system of equations:
\begin{equation}
\begin{split}
& N_1 = R_{\ttbar} N_{1,\ttbar} + R_{\PW} N_{1,{\PW}} + N_{1,QCD}  +  N_{1,O}; \\
& N_2 = R_{\ttbar} N_{2,\ttbar} + R_{\PW} N_{2,{\PW}} + N_{2,QCD}  +  N_{2,O}.
\end{split}
\end{equation}
where $N_{i}$, $N_{i,{\PW}}$, $N_{i,O}$,
$N_{i,\ttbar}$, and  $N_{i,QCD}$
are the number of events in data, \PW+jets, other MC backgrounds, $\ttbar$-MC,
and QCD multijet events obtained from data, passing selection $i$. In the $\Pe \nu jj$ analysis, the solution of the system yields the following normalization factors: $R_{\ttbar} = 0.82 \pm 0.04$~(stat.)$\pm 0.02$~(syst.) and $R_{\PW} = 1.21 \pm 0.03\,$(stat.)$\pm 0.02\,$(syst.), where the systematic errors reflect the uncertainties on the QCD multijet background estimate. In the $\mu \nu jj$ analysis, the normalization factors are: $R_{\ttbar} = 0.84 \pm 0.03$~(stat.) and $R_{\PW} = 1.24 \pm 0.02\,$(stat.).

The contribution from QCD multijet processes after all selection criteria are
applied to the $\mu\nu jj$ sample is estimated to be negligible. No QCD MC events survive the full selection
criteria optimized for any leptoquark mass hypothesis. However, as multijet
processes are difficult to accurately model by MC simulation,
several crosschecks are made with data control samples to ensure that the QCD multijet
background in the $\mu\nu jj$ analysis is negligible. The method used to determine the QCD multijet background in the $\Pe \nu jj$ analysis is similar to the one used for the $\Pe\Pe jj$ channel.

\section{Event selection optimization}
\label{selection}
After the initial selection, the sensitivity of the search is optimized by maximizing the Gaussian signal significance $S/\sqrt{S+B}$ in all channels. Optimized thresholds on the following variables are applied for each leptoquark mass hypothesis in the $\ell\ell jj$ channels: $M_{\ell\ell}$, $\ST^{\ell\ell}$, and $M^{\text{min}}_{\ell j}$. The invariant mass of the dilepton pair, $M_{\ell\ell}$, is used to remove the majority of the contribution from the  $\cPZ/\gamma^*$+jets background.
The variable $M^{\text{min}}_{\ell j}$ is defined as the smaller lepton-jet invariant mass for the assignment of jets and leptons to leptoquarks which minimizes the LQ~-~$\overline{\mathrm {LQ}}$ invariant mass difference.

Thresholds on the following variables are optimized for each leptoquark mass hypothesis in the $\ell\nu jj$ channels: $\MET$, $\ST^{\ell\nu}$, and $M_{\ell j}$. A minimum threshold on $\MET$ is used, primarily to reduce the dominant \PW+jets background. The variable $M_{\ell j}$ is defined as the invariant mass of the lepton-jet combination which minimizes the  LQ~-~$\overline{\mathrm {LQ}}$ transverse mass difference. In addition, a lower threshold is applied on the transverse mass of the lepton (electron or muon) and $\MET$ in the event, $\MT>120$\GeV.

The resulting optimized thresholds %for {2.0~fb$^{-1}$} of integrated luminosity
are summarized in Tables~\ref{tab:optimization_ll} and~\ref{tab:optimization_lnu} for the $\ell\ell jj$ and the $\ell\nu jj$ channels, respectively.

\begin{table}[htb]
\topcaption{Optimized thresholds for different mass hypothesis of the $\ell\ell jj$ signal.}
\label{tab:optimization_ll}
\scriptsize
\begin{center}
\begin{tabular}{|lccccccccccc|}
\hline
$M_{\mathrm{LQ}}$ (\GeVns{}) & 250 & 350 & 400 & 450 & 500 & 550 & 600 & 650 & 750 & 850 & 900\\ \hline\hline
$\ST^{\ell\ell} > $  (\GeVns{}) &  330 &  450 &  530 &  610 &  690 &  720 &  770 &  810 &  880 &  900  & 920 \\
$M_{\ell\ell} >$ (\GeVns{}) &  100 &  110 &  120 &  130 &  130 &  130 &  130 &  130 &  140 &  150  & 150 \\
$ M^{\text{min}}_{\ell j} >$ (\GeVns{}) &  60 &  160 &  200 &  250 &  300 &  340 &  370 &  400 &  470 &  500 & 520 \\ \hline
\end{tabular}
\end{center}
\end{table}

\begin{table} [htb]
     \topcaption{Optimized thresholds for different mass hypotheses of the $\ell\nu jj$ signal.}
     \label{tab:optimization_lnu}
\scriptsize
 \begin{center}
\begin{tabular}{| l  c  c  c  c  c  c  c  c  c  c | }
  \hline
$M_{\mathrm{LQ}}$ (\GeVns{})  & 250 & 350 & 400 & 450 & 500 & 550 & 600 & 650 & 750 & 850 \\ \hline\hline
$\ST^{\ell\nu}>$ (\GeVns{})  & 450 & 570 & 650 & 700 & 800 & 850 & 890 & 970 & 1000 & 1000 \\
$\MET >$ (\GeVns{})  & 100 & 120 & 120 & 140 & 160 & 160 & 180 & 180 & 220 & 240 \\
$ M_{\ell j} >$  (\GeVns{})  & 150 & 300 & 360 & 360 & 360 & 480 & 480 & 540 & 540 & 540 \\ \hline
\end{tabular}
\end{center}
 \end{table}

After the initial selection criteria are applied, the yields in data are found to be consistent with SM predictions. Distributions of variables used in the final selection for the $\Pe\Pe jj$, $\Pe \nu jj$, $\mu\mu jj$, and $\mu\nu jj$ analyses are shown in Figs.~\ref{figapp:stwt_ee}--\ref{figapp:stwt_munu}.

\begin{figure}[htbp]
       \begin{center}
       {\includegraphics[width=.48\textwidth]{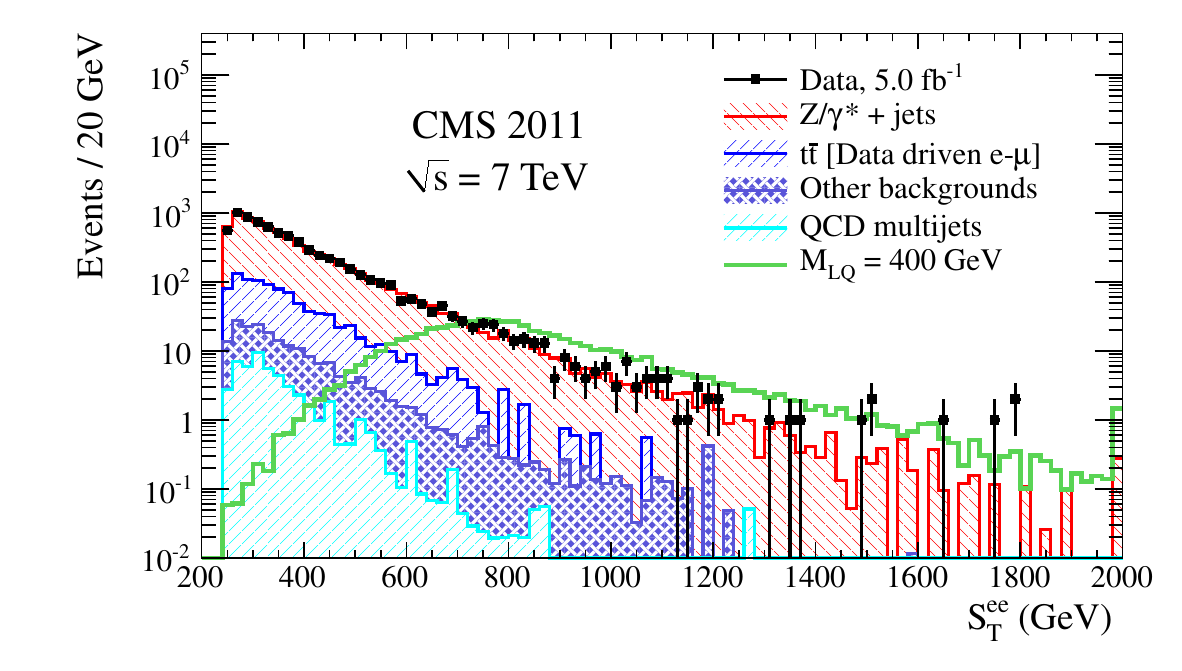}}
       {\includegraphics[width=.48\textwidth]{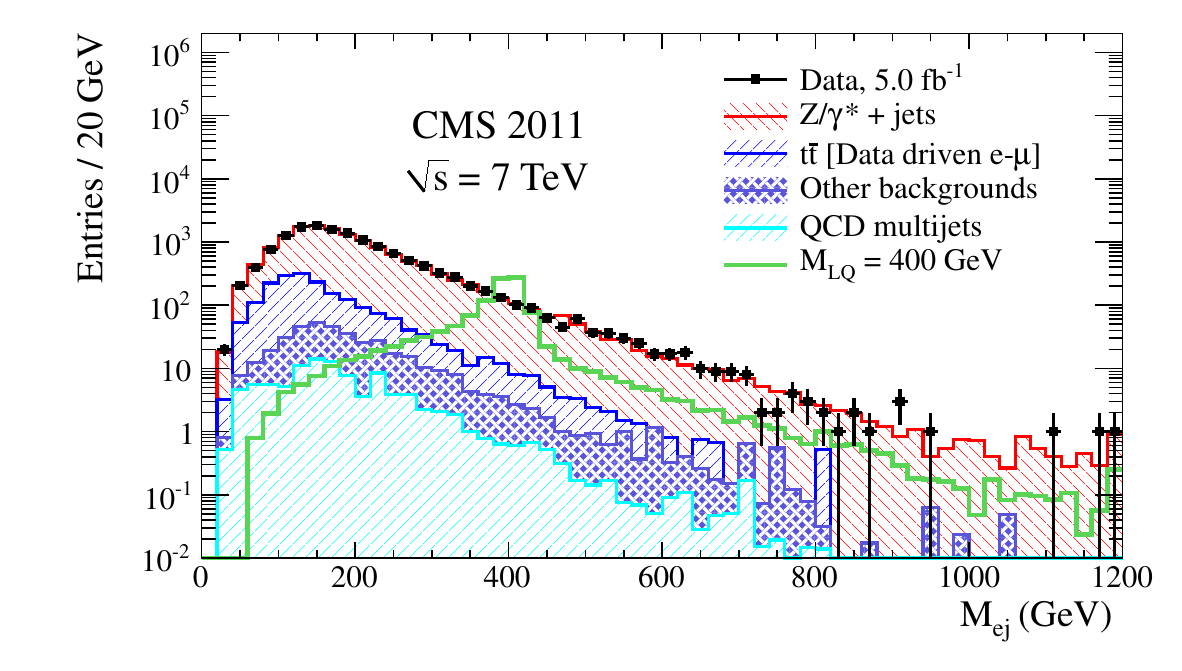}}
       \caption{$\Pe\Pe jj$ channel: the distributions of $\ST^{{\Pe\Pe}}$ (\cmsLeft) and of $ M_{\Pe j}$ for each of the two electron-jet pairs (\cmsRight) for events that pass the initial selection level. The data are indicated by the points, and the SM backgrounds are given as cumulative histograms. The expected contribution from a leptoquark signal with $M_{\mathrm {LQ}}=400$\GeV is also shown.}
	  \label{figapp:stwt_ee}
\end{center}
\end{figure}

\begin{figure}[htbp!]
       \begin{center}
       {\includegraphics[width=.48\textwidth]{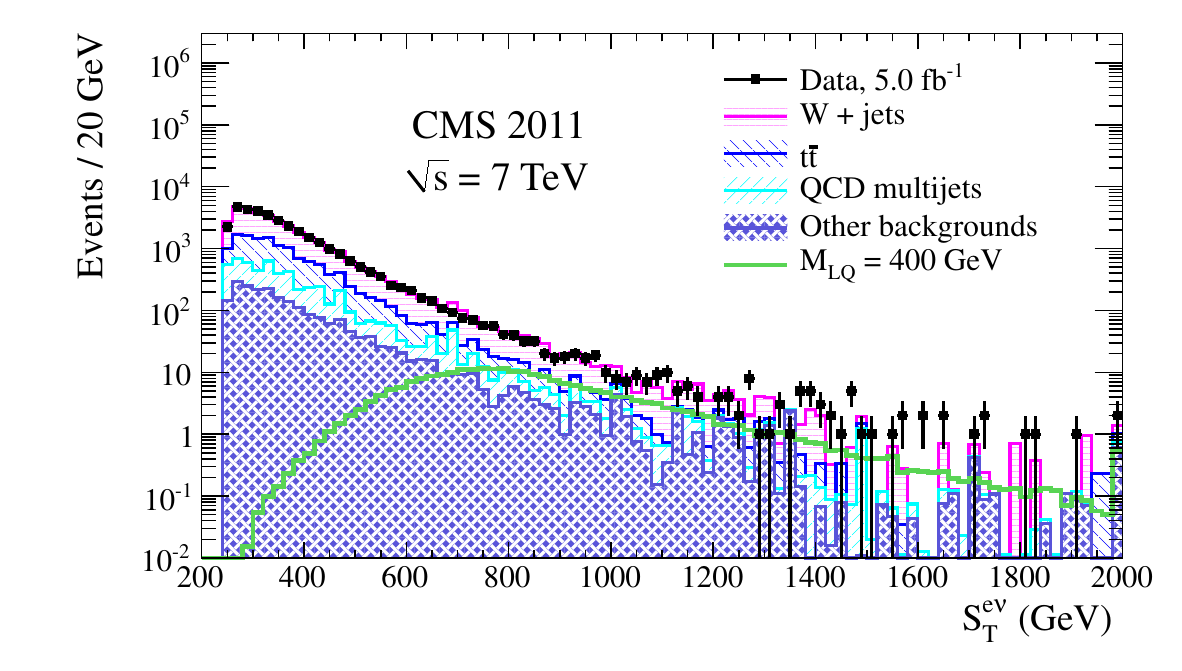}}
       {\includegraphics[width=.48\textwidth]{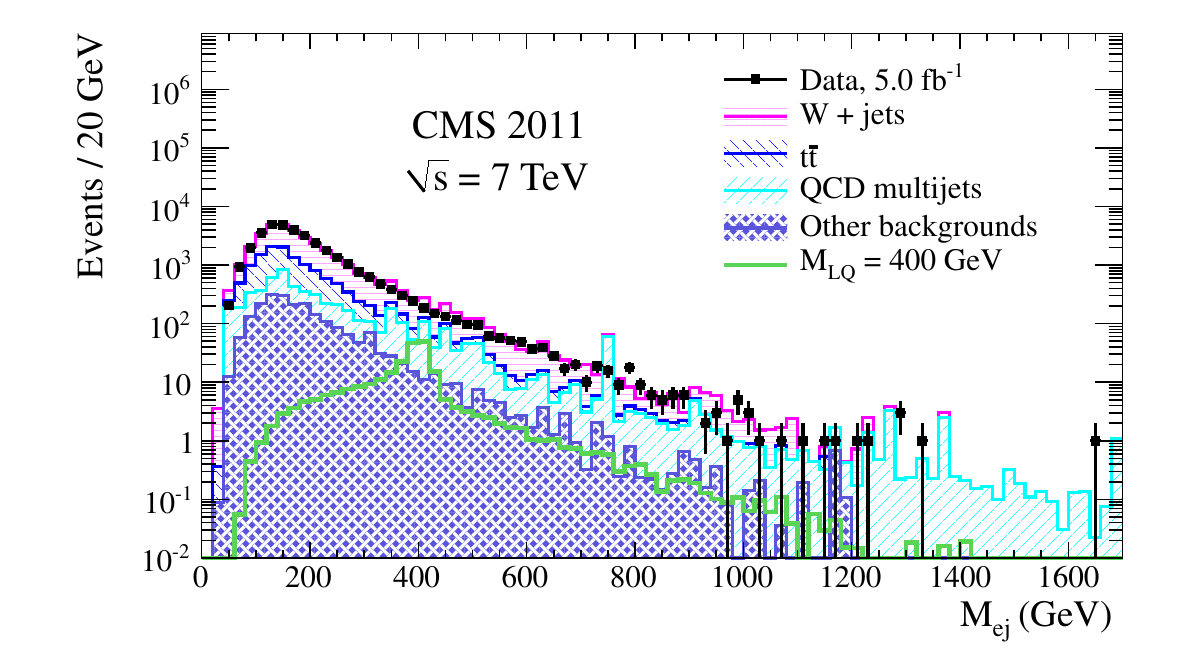}}
       \caption{$\Pe\nu jj$ channel: the distributions of $\ST^{\Pe\nu}$ (\cmsLeft) and of $M_{\Pe j}$ (\cmsRight) for events that pass the initial selection level. The data are indicated by the points, and the SM backgrounds are given as cumulative histograms. The expected contribution from a leptoquark signal with $M_{\mathrm {LQ}}=400$\GeV is also shown.}
	  \label{figapp:stwt_enu}
      \end{center}
\end{figure}

\begin{figure}[htbp!]
       \begin{center}
       {\includegraphics[width=.48\textwidth]{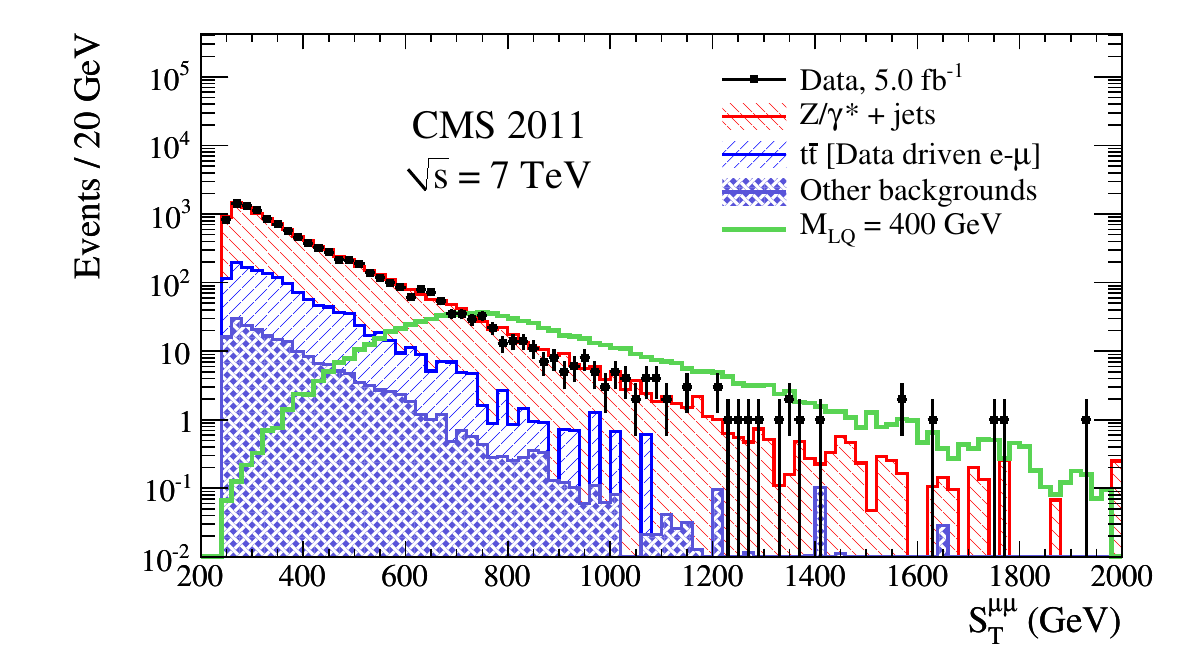}}
       {\includegraphics[width=.48\textwidth]{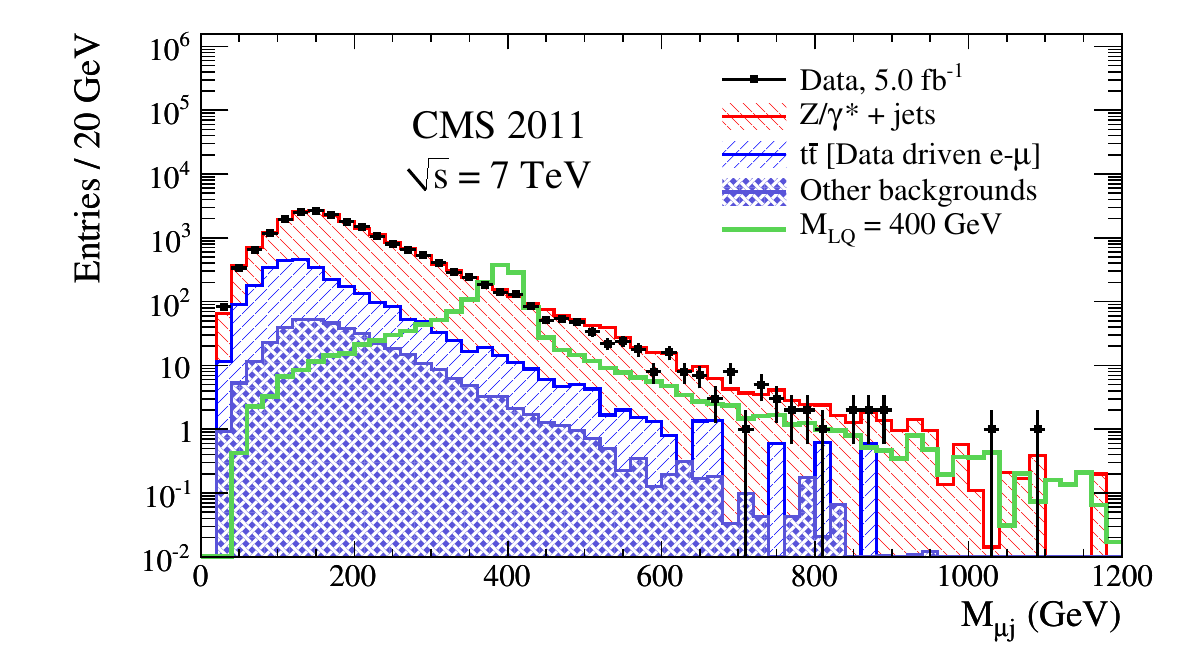}}
       \caption{$\mu\mu jj$ channel: the distributions of $\ST^{\mu\mu}$ (\cmsLeft) and of $ M_{\mu j}$ for each of the two muon-jet pairs (\cmsRight) for events that pass the initial selection level. The data are indicated by the points, and the SM backgrounds are given as cumulative histograms. The expected contribution from a leptoquark signal with $M_{\mathrm {LQ}}=400$\GeV is also shown.}
	  \label{figapp:stwt_mumu}
      \end{center}
\end{figure}

\begin{figure}[htbp!]
       \begin{center}
       {\includegraphics[width=.48\textwidth]{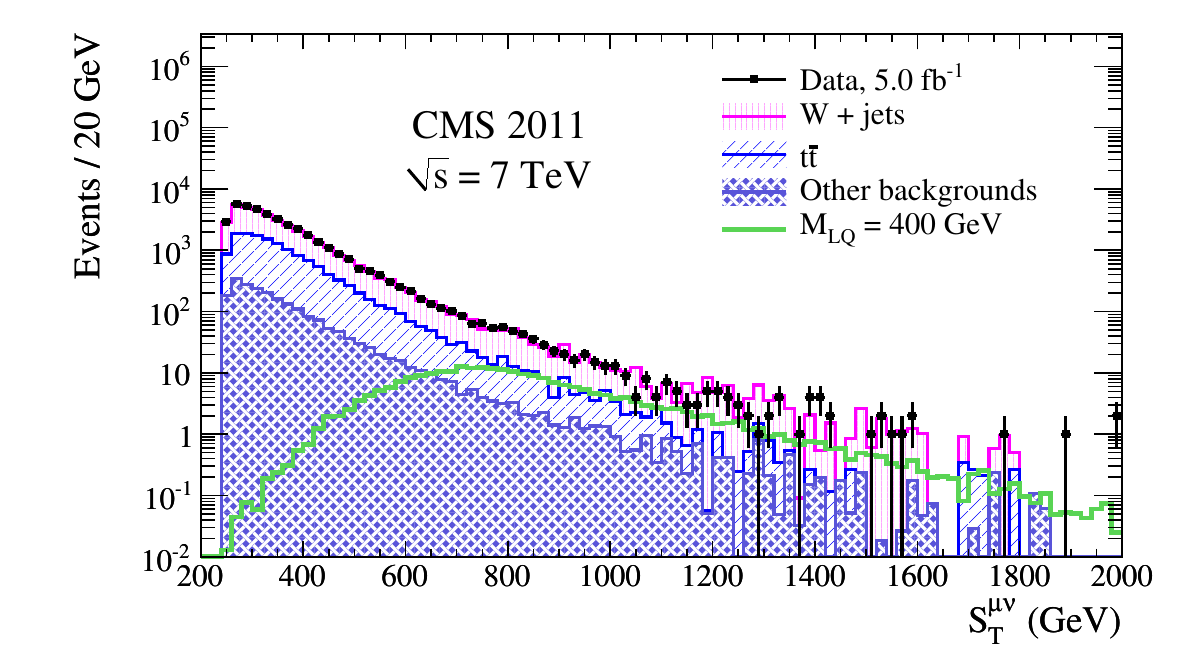}}
       {\includegraphics[width=.48\textwidth]{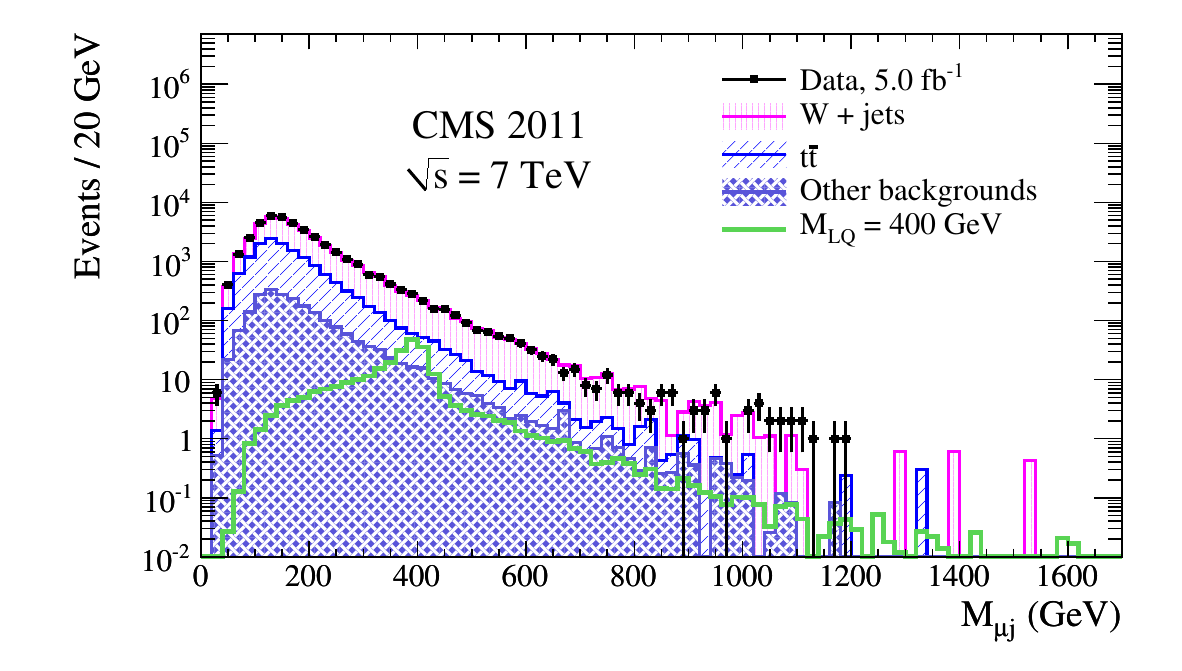}}
       \caption{$\mu\nu jj$ channel: the distributions of $\ST^{\mu\nu}$ (\cmsLeft) and of $M_{\mu j}$ (\cmsRight) for events that pass the initial selection level. The data are indicated by the points, and the SM backgrounds are given as cumulative histograms. The expected contribution from a leptoquark signal with $M_{\mathrm {LQ}}=400$\GeV is also shown.}
	  \label{figapp:stwt_munu}
      \end{center}
\end{figure}

The number of events selected in data and estimated backgrounds are then compared at different stages of selection. This information is shown in Tables~\ref{tab:eejjFinalSelection}--\ref{tab:finalselection_munu} for the initial selection and for the final selection for each channel separately.

    \begin{table*}[htb]
\topcaption{Individual background (BG) sources, expected signal, data and total background event yields after the initial (first row) and final selections for the $\Pe\Pe jj$ analysis. Other BG includes single top,\PW+jets, $\gamma$+jets, and $VV$+jets.  Only statistical uncertainties are reported.}
    \label{tab:eejjFinalSelection}
    \begin{center}
    \footnotesize
\begin{tabular}{| l  c  c  c  c  c  c  c |}
  \hline
$M_{\mathrm{LQ}}$  & \cPZ+jets & $\ttbar$ & QCD & Other BG & LQ Signal& Data &  Total BG \\
  \hline
  \hline
--                  &  $ 6234 \pm 24 $    & $ 768 \pm 19 $           & $ 49.59 \pm 0.43 $       & $ 147.6 \pm 2.3 $   &  -- & 7201 & $ 7199 \pm 31 $         \\
  [2pt]\hline
400 &    $ 35.7 \pm 1.8 $   & $ 19.1 \pm 3.1 $         &  $ 0.877 \pm 0.022 $     & $ 3.12 \pm 0.56 $   &$ 487.4 \pm 2.2 $    & 55  & $ 58.8 \pm 3.6 $        \\[2pt]
500 &    $ 6.55 \pm 0.70 $  & $ 2.45 \pm 1.10 $        &  $ 0.192 \pm 0.012 $     & $ 1.03 \pm 0.42 $   &$ 109.30 \pm 0.46 $  & 14  & $ 10.2 \pm 1.4 $        \\[2pt]
550 &    $ 4.65 \pm 0.58 $  & $ 0.98 \pm 0.69 $        &  $ 0.139 \pm 0.012 $     & $ 0.84 \pm 0.42 $   &$ 57.35 \pm 0.23 $   & 11  & $ 6.60 \pm 0.99 $       \\[2pt]
600 &    $ 3.04 \pm 0.46 $  & $ 0.49 \pm 0.49 $        &  $ 0.088 \pm 0.011 $     & $ 0.72 \pm 0.41 $   &$ 30.95 \pm 0.14 $   & 8   & $ 4.34 \pm 0.79 $       \\[2pt]
650 &    $ 2.14 \pm 0.38 $  & $ 0.49 \pm 0.49 $        &  $ 0.073 \pm 0.011 $     & $ 0.48 \pm 0.40 $   &$ 16.998 \pm 0.065 $ & 6   & $ 3.18 \pm 0.74 $       \\[2pt]
750 &    $ 1.04 \pm 0.26 $  & $ 0.000_{-0.00}^{+0.56}$ &  $ 0.0092 \pm 0.0020 $   & $ 0.41 \pm 0.40 $   &$ 5.526 \pm 0.023 $  & 0   & $ 1.45_{-0.47}^{+0.73}$ \\[2pt]
850 &   $ 0.81 \pm 0.23 $  & $ 0.000_{-0.00}^{+0.56}$ &  $ 0.00101 \pm 0.00022 $ & $ 0.40 \pm 0.40 $   &$ 1.9679 \pm 0.0078 $ & 0   & $ 1.21_{-0.46}^{+0.72}$ \\[2pt]
  \hline
\end{tabular}
    \end{center}
    \end{table*}

    \begin{table*}[htb]
\topcaption{Individual background (BG) sources, expected signal, data, and total background event yields after the initial (first row) and final selections for the $\Pe \nu jj$ analysis.  Other BG includes single top, \cPZ+jets, $\gamma$+jets, and $VV$+jets.  Only statistical uncertainties are reported.}
    \label{tab:enujjFinalSelection}
    \begin{center}
    \footnotesize
\begin{tabular}{| l  c  c  c  c  c  c  c |}
  \hline
$M_{\mathrm{LQ}}$  & \PW+jets & $\ttbar$ & QCD & Other& LQ Signal & Data &  Total BG \\
  \hline
  \hline
-- &   $ 20108   \pm 99 $   & $ 9301   \pm 42   $ & $ 3267    \pm 26    $  & $ 1913 \pm 53 $& -- &34135 & $ 34590 \pm 120 $ \\
  [2pt]\hline
400 &    $ 28.7 \pm 3.6 $   & $ 17.5 \pm 1.8 $    & $ 6.20 \pm 0.46 $    & $ 6.01 \pm 0.77 $ &$ 126.01 \pm 0.82 $   & 43  & $ 58.4 \pm 4.1 $ \\[2pt]
500 &    $ 13.3 \pm 2.4 $   & $ 6.3 \pm 1.1 $     & $ 1.72 \pm 0.22 $    & $ 2.80 \pm 0.37 $ &$ 34.70 \pm 0.23 $    & 18  & $ 24.2 \pm 2.6 $ \\[2pt]
550 &    $ 2.98 \pm 0.95 $  & $ 3.38 \pm 0.82 $   & $ 0.65 \pm 0.10 $    & $ 1.46 \pm 0.26 $ &$ 16.25 \pm 0.10 $    & 10  & $ 8.5 \pm 1.3 $ \\[2pt]
600 &    $ 2.45 \pm 0.87 $  & $ 2.33 \pm 0.67 $   & $ 0.57 \pm 0.10 $    & $ 1.29 \pm 0.25 $ &$ 9.442 \pm 0.056 $   & 6   & $ 6.6 \pm 1.1 $ \\[2pt]
650 &    $ 2.03 \pm 0.83 $  & $ 1.01 \pm 0.41 $   & $ 0.335 \pm 0.079 $  & $ 0.76 \pm 0.20 $ &$ 5.202 \pm 0.032 $   & 4   & $ 4.14 \pm 0.95 $ \\[2pt]
750 &    $ 1.45 \pm 0.65 $  & $ 0.62 \pm 0.31 $   & $ 0.287 \pm 0.080 $  & $ 0.65 \pm 0.18 $ &$ 1.851 \pm 0.010 $   & 4   & $ 3.01 \pm 0.75 $ \\[2pt]
850 &    $ 1.22 \pm 0.61 $  & $ 0.62 \pm 0.31 $   & $ 0.251 \pm 0.078 $  & $ 0.61 \pm 0.19 $ &$ 0.6973 \pm 0.0037 $ & 4   & $ 2.70 \pm 0.71 $ \\[2pt]
  \hline
\end{tabular}
     \end{center}
    \end{table*}

\begin{table*}[htb]
\topcaption{Individual background (BG) sources, expected signal, data and total background event yields after the initial (first row) and final selections for the $\mu\mu jj$ analysis.  Other BG includes single top, \PW+jets, and $VV$+jets. Only statistical uncertainties are reported.}
\begin{center}
\begin{tabular}{|lcccccc|}\hline
 $M_{\mathrm {LQ}}$    & \cPZ+jets               & $\ttbar$                    & Other BG                      & LQ Signal             & Data        & Total BG                   \\ \hline \hline
 --          & $ 8644 \pm 47 $      & $ 1218 \pm 27 $             & $ 201.7 \pm 2.8 $             &  --                   & $ 9897 $    & $ 10063 \pm 54 $            \\[2pt]\hline
 400         & $ 46.9 \pm 2.4 $     & $ 30.1 \pm 4.2 $            & $ 3.58 ^{+0.70}_{-0.40} $     & $ 629.3 \pm 4.0 $     & $ 68 $      & $ 80.6 ^{+4.9}_{-4.9} $    \\[2pt]
 500         & $ 10.4 \pm 1.1 $     & $ 4.1 \pm 1.6 $             & $ 0.89 ^{+0.62}_{-0.21} $     & $ 136.73 \pm 0.86 $   & $ 14 $      & $ 15.4 ^{+2.0}_{-1.9} $    \\[2pt]
 550         & $ 7.29 \pm 0.94 $    & $ 2.4 \pm 1.2 $             & $ 0.53 ^{+0.60}_{-0.17} $     & $ 70.49 \pm 0.40 $    & $ 9 $       & $ 10.2 ^{+1.6}_{-1.5} $     \\[2pt]
 600         & $ 5.03 \pm 0.75 $    & $ 0.59 \pm 0.59 $           & $ 0.47 ^{+0.60}_{-0.16} $     & $ 37.39 \pm 0.22 $    & $ 6 $       & $ 6.1 ^{+1.1}_{-1.0} $     \\[2pt]
 650         & $ 3.82 \pm 0.65 $    & $ 0.59 \pm 0.59 $           & $ 0.24 ^{+0.59}_{-0.12} $     & $ 20.56 \pm 0.13 $    & $ 5 $       & $ 4.7 ^{+1.1}_{-0.9} $     \\[2pt]
 750         & $ 2.03 \pm 0.47 $    & $ 0.00 ^{+0.67}_{-0.00} $   & $ 0.09 ^{+0.59}_{-0.08} $     & $ 6.529 \pm 0.038 $   & $ 1 $       & $ 2.1 ^{+1.0}_{-0.5} $     \\[2pt]
 850         & $ 1.56 \pm 0.42 $    & $ 0.00 ^{+0.67}_{-0.00} $   & $ 0.08 ^{+0.59}_{-0.08} $     & $ 2.327 \pm 0.014 $   & $ 0 $       & $ 1.6^{+1.0}_{-0.4} $      \\[2pt] \hline
\end{tabular}
\label{tab:finalselection_mumu}
\end{center}
\end{table*}

\begin{table*}[htbp]
\topcaption{Individual background (BG) sources, expected signal, data, and total background event yields after the initial (first row) and final selections for the $\mu\nu jj$ analysis.  Other BG includes single top, \cPZ+jets, and $VV$+jets.  Only statistical uncertainties are reported.}
\begin{center}
\begin{tabular}{|lcccccc|}\hline
 $M_{\mathrm {LQ}}$     & \PW+jets             & $\ttbar$             & Other BG           & LQ Signal              & Data        & Total BG            \\ \hline \hline
 --           & $ 23910 \pm 161 $   & $ 12254 \pm 58 $     & $ 2146 \pm 18 $    & --                     & $ 39585 $   & $ 38311 \pm 173 $   \\[2pt]\hline
 400          & $ 31.8 \pm 4.4 $    & $ 23.9 \pm 2.6 $     & $ 5.68 \pm 0.64 $  & $ 118.3 \pm 1.2 $       & $ 60 $      & $ 61.5 \pm 5.2 $    \\[2pt]
 500          & $ 18.0 \pm 3.3 $    & $ 9.6 \pm 1.7 $      & $ 3.38 \pm 0.51 $  & $ 33.94 \pm 0.28 $     & $ 26 $      & $ 31.0 \pm 3.7 $    \\[2pt]
 550          & $ 5.7 \pm 1.8 $     & $ 4.2 \pm 1.1 $      & $ 2.21 \pm 0.43 $  & $ 15.51 \pm 0.14  $    & $ 12 $      & $ 12.1 \pm 2.1 $    \\[2pt]
 600          & $ 5.5 \pm 1.8 $     & $ 3.2 \pm 1.0 $      & $ 1.80 \pm 0.39 $  & $ 9.23 \pm 0.08 $      & $ 8 $       & $ 10.5 \pm 2.1 $    \\[2pt]
 650          & $ 2.9 \pm 1.2 $     & $ 1.14 \pm 0.59 $    & $ 1.13 \pm 0.31 $  & $ 4.964 \pm 0.043 $    & $ 7 $       & $ 5.1 \pm 1.3 $     \\[2pt]
 750          & $ 2.9 \pm 1.2 $     & $ 0.51 \pm 0.36 $    & $ 0.99 \pm 0.29 $  & $ 1.840 \pm 0.015 $     & $ 6 $       & $ 4.4 \pm 1.2 $     \\[2pt]
 850          & $ 2.7 \pm 1.1 $     & $ 0.51 \pm 0.36 $    & $ 0.76 \pm 0.23 $  & $ 0.6906 \pm 0.0051 $  & $ 6 $       & $ 4.0 \pm 1.2 $     \\[2pt] \hline
\end{tabular}
\label{tab:finalselection_munu}
\end{center}
\end{table*}

Data and background predictions after final selection are also shown in Figs.~\ref{figapp:stmlq600_ee}--\ref{figapp:stmlq600_munu}, which compare $\ST$ and the best combination for the lepton-jet invariant mass for a signal leptoquark mass of 600\GeV in the four decay channels considered.

\begin{figure}[htbp]
       \begin{center}
       {\includegraphics[width=.48\textwidth]{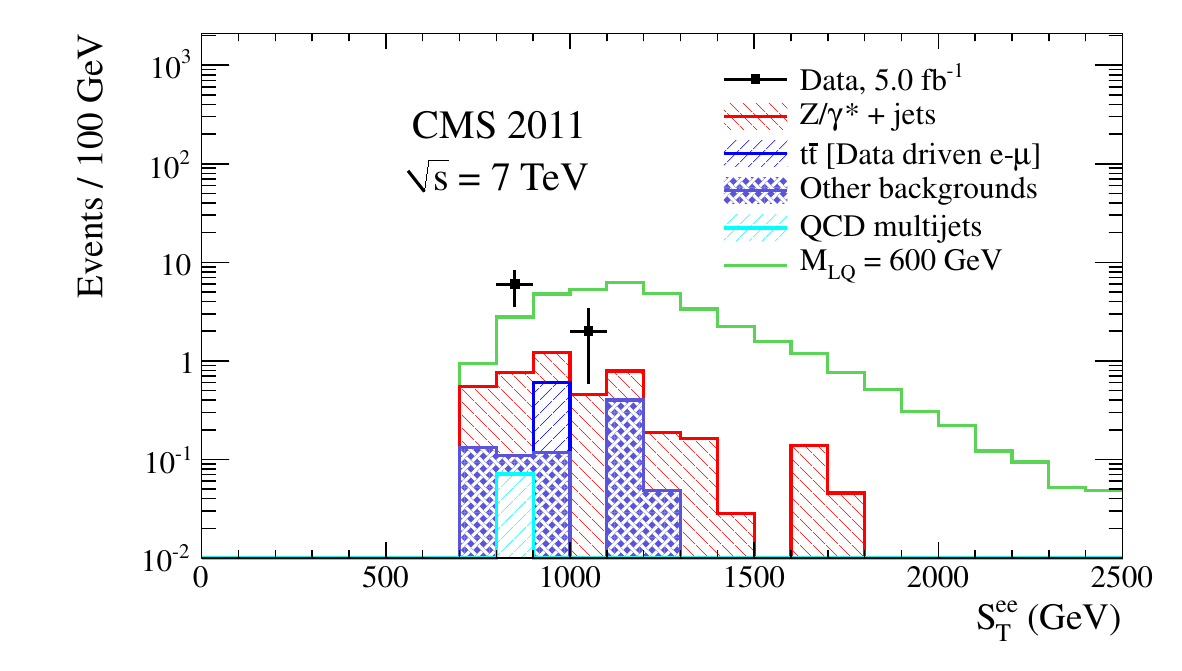}}
       {\includegraphics[width=.48\textwidth]{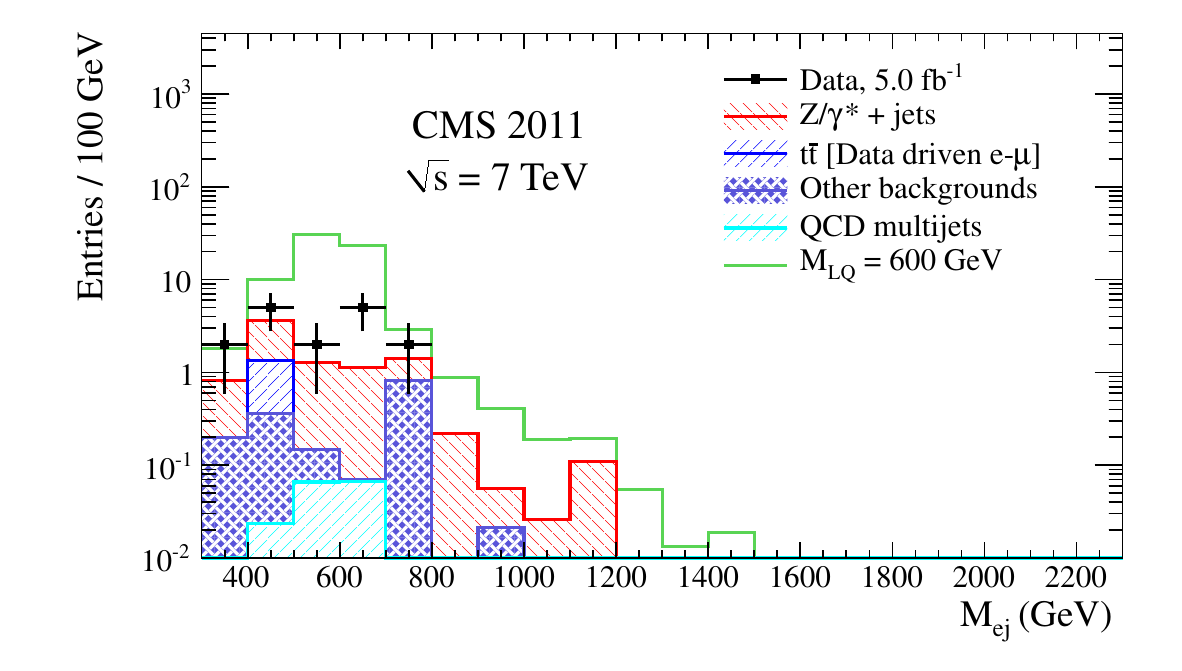}}
       \caption{$\Pe\Pe jj$ channel: the distributions of $\ST^{\Pe\Pe}$ (\cmsLeft) and of $ M_{\Pe j}$ for each of the two electron-jet pairs (\cmsRight) for events that pass the final selection criteria optimized for a signal leptoquark mass of 600\GeV. The data are indicated by the points, and the SM backgrounds are given as cumulative histograms. The expected contribution from a leptoquark signal with $M_{\mathrm {LQ}}=600$\GeV is also shown.}
	  \label{figapp:stmlq600_ee}
	  \end{center}
\end{figure}

\begin{figure*}[htbp!]
       \begin{center}
       {\includegraphics[width=.48\textwidth]{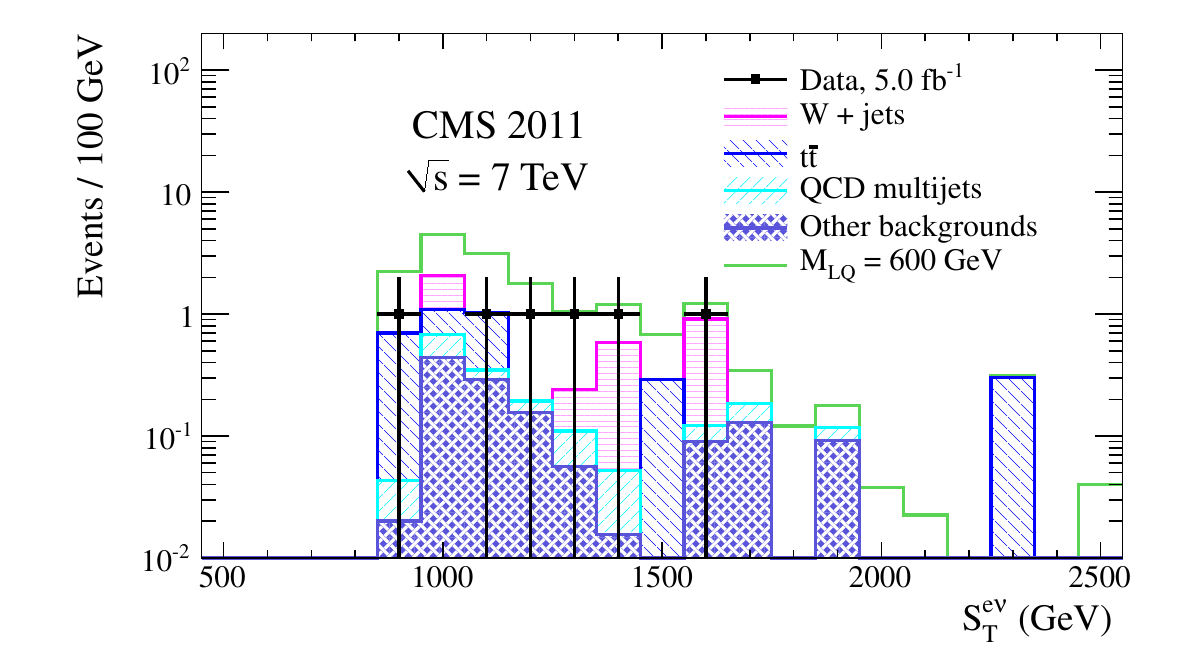}}
       {\includegraphics[width=.48\textwidth]{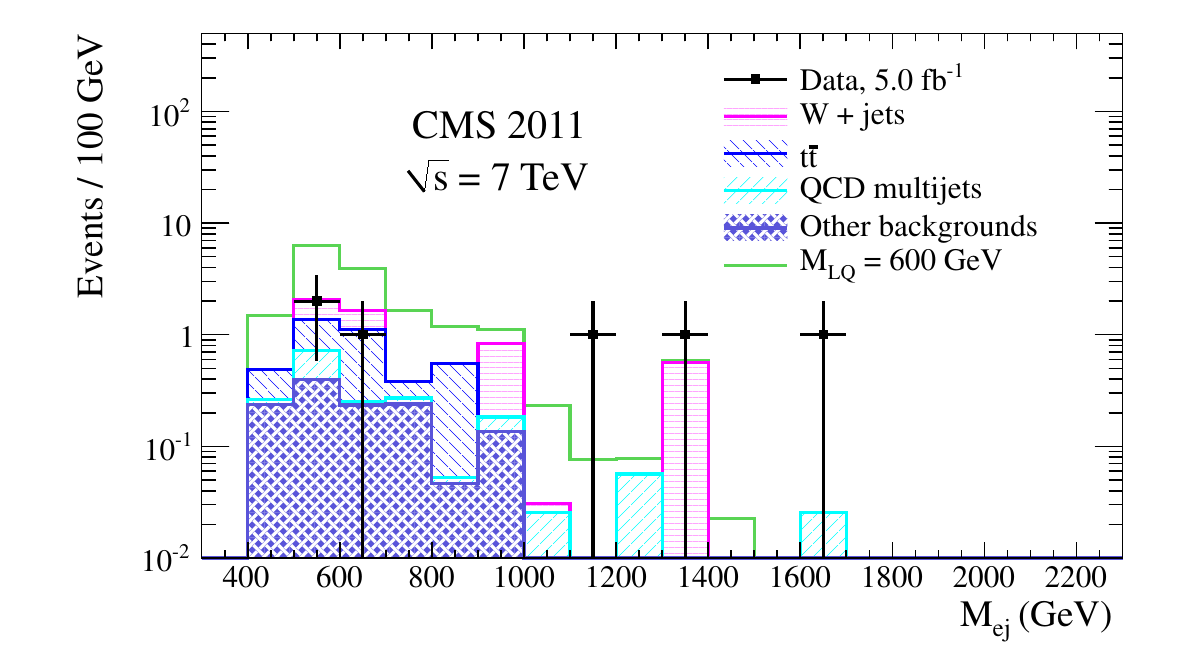}}
       \caption{$\Pe\nu jj$ channel: the distributions of $\ST^{\Pe\nu}$ (left) and of $M_{\Pe j}$ (right) for events that pass the final selection criteria optimized for a signal leptoquark mass of 600\GeV. The data are indicated by the points, and the SM backgrounds are given as cumulative histograms. The expected contribution from a leptoquark signal with $M_{\mathrm {LQ}}=600$\GeV is also shown.}
	  \label{figapp:stmlq600_enu}
	  \end{center}
\end{figure*}

\begin{figure*}[htbp!]
       \begin{center}
       {\includegraphics[width=.48\textwidth]{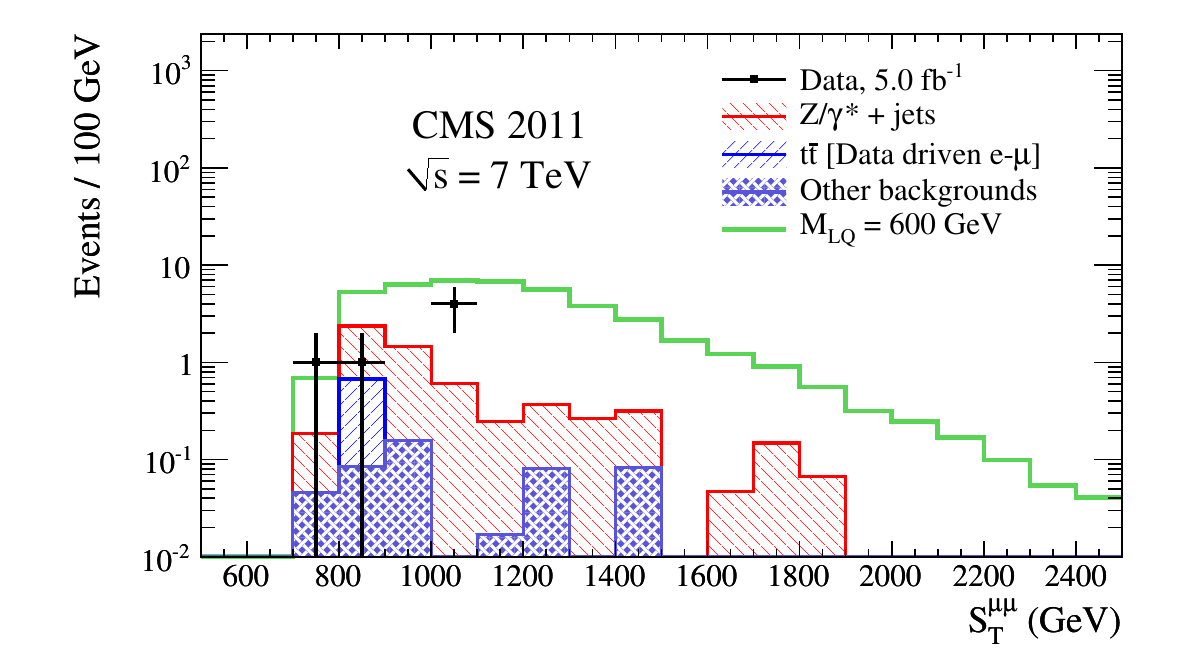}}
       {\includegraphics[width=.48\textwidth]{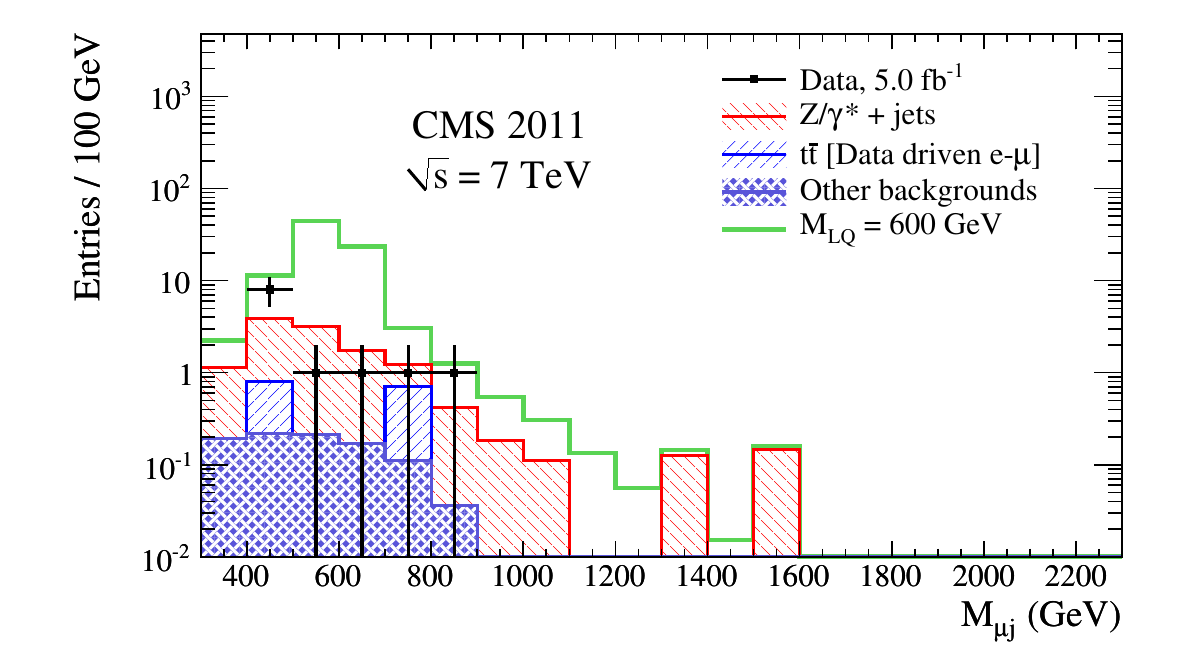}}
       \caption{$\mu\mu jj$ channel: the distributions of $\ST^{\mu\mu}$ (left) and of $ M_{\mu j}$ for each of the two muon-jet pairs (right) for events that pass the final selection criteria optimized for a signal leptoquark mass of 600\GeV. The data are indicated by the points, and the SM backgrounds are given as cumulative histograms. The expected contribution from a leptoquark signal with $M_{\mathrm {LQ}}=600$\GeV is also shown.}
	  \label{figapp:stmlq600_mumu}
	  \end{center}
\end{figure*}

\begin{figure*}[htbp!]
       \begin{center}
       {\includegraphics[width=.48\textwidth]{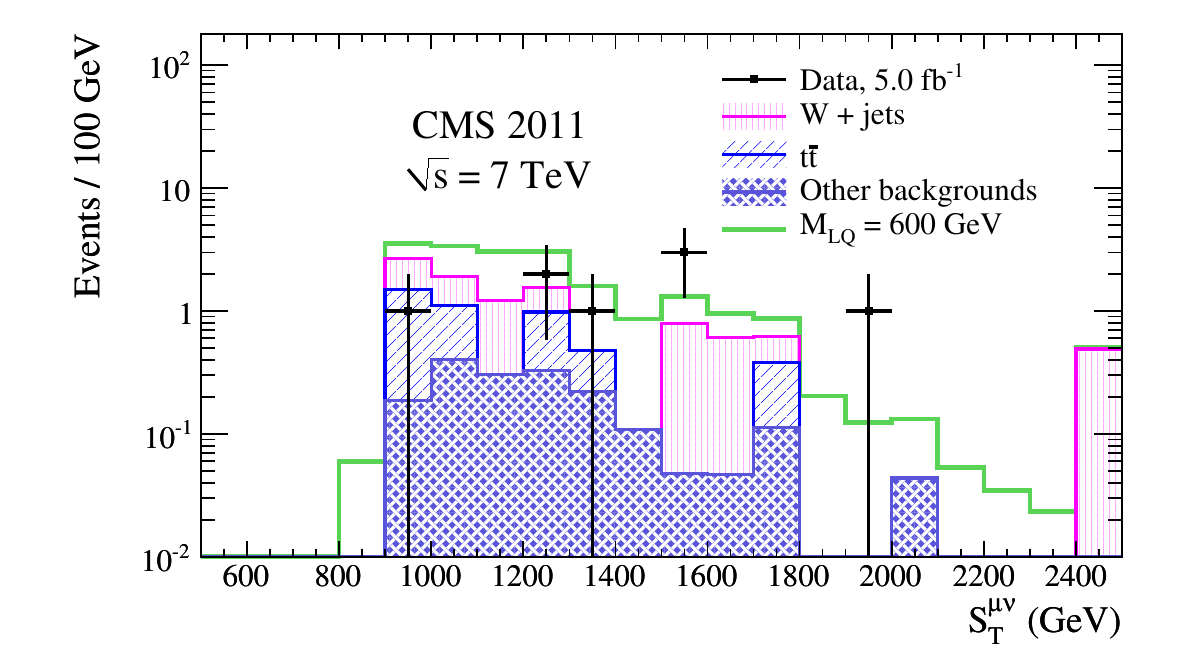}}
       {\includegraphics[width=.48\textwidth]{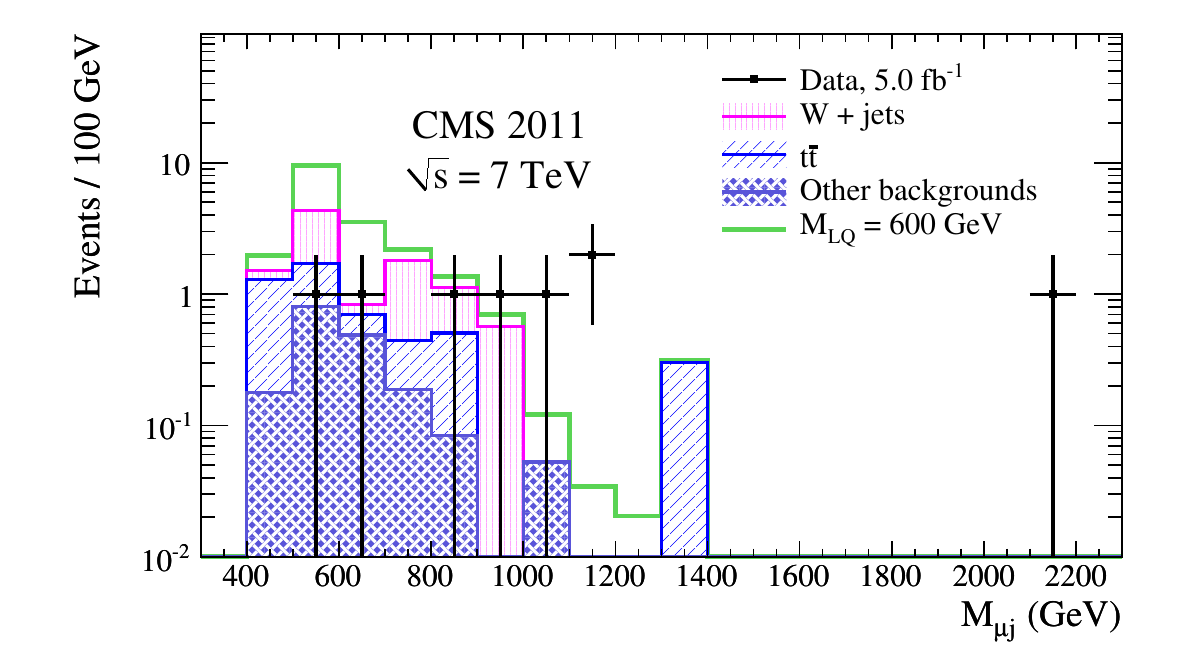}}
       \caption{$\mu\nu jj$ channel: the distributions of $\ST^{\mu\nu}$ (left) and of $M_{\mu j}$ (right) for events that pass the final selection criteria optimized for a signal leptoquark mass of 600\GeV. The data are indicated by the points, and the SM backgrounds are given as cumulative histograms. The expected contribution from a leptoquark signal with $M_{\mathrm {LQ}}=600$\GeV is also shown.}
	  \label{figapp:stmlq600_munu}
	  \end{center}
\end{figure*}

\section{Systematic Uncertainties}
\label{systematics}
The uncertainty on the integrated luminosity is taken as 2.2\%~\cite{lumi_new}.

The statistical uncertainties on the values of $R_{\cPZ}~(R_{\PW})$ after initial selection requirements are used as an estimate of the uncertainty on the normalization of
the $\cPZ/\gamma^*$+jets~(W+jets) backgrounds.
The uncertainty on
the shape of the $\cPZ/\gamma^*$+jets and \PW+jets distributions is calculated to be $15\%~(10\%)$ and $20\%~(11\%)$ for first (second) generation, respectively, by comparing the predictions of {\sc MadGraph} samples produced with factorization or renormalization
scales and  matrix element--parton shower matching threshold varied up and down by a factor of two.

The uncertainty on the estimate of the $\ttbar$ background in the
$\Pe\Pe jj$ and $\mu\mu jj$ channels is derived from the statistical uncertainty of the $\Pe \mu jj$ data sample and the ratio of electron and muon reconstruction
uncertainties, which is calculated to be $2\%$ and $3\%$, respectively. In addition, an uncertainty of $7\%$ is assigned based on the estimated contamination from sources other than $\ttbar$ in the data sample containing one electron and one muon.  A $5.5\%~(5\%)$ uncertainty on the normalization of the estimated $\ttbar$ background in the $\Pe \nu jj~(\mu\nu jj)$ channel is given by the statistical uncertainty on the value of $R_{\ttbar}$ after the initial selection requirements.
A $10\%~(10\%)$ uncertainty on the shape of the $\ttbar$ background distribution in the $\Pe \nu jj~(\mu\nu jj)$ channel is estimated by comparing the predictions of \MADGRAPH samples produced with factorization or renormalization scales and  matrix element--parton shower matching thresholds varied up and down by a factor of two.

A systematic uncertainty of 50\% (25\%) on the QCD multijet background estimate for the $\Pe\Pe jj$ ($\Pe\nu jj$) channel is estimated from the difference between the number of observed data and the background prediction in a QCD multijet-enriched data sample with lower jet multiplicity.

PDF uncertainties on the theoretical cross section of leptoquark production and on the final selection acceptance have been calculated using the PDF4LHC ~\cite{pdf4lhc} prescriptions, with PDF and $\alpha_s$ variations of the MSTW2008~\cite{Martin:2009iq}, CTEQ6.6~\cite{Nadolsky:2008zw} and NNPDF2.0~\cite{Ball:2010de} PDF sets taken into account. Uncertainties on the cross section vary from 10 to 30 \% for leptoquarks in the mass range of 200--900\GeV, while the effect of the PDF uncertainties on signal acceptance varies from 1 to 3\%. The PDF uncertainties are not considered for background sources with uncertainties determined from data. An uncertainty on the modeling of pileup interactions in the MC simulation is determined by varying the mean of the distribution of pileup interactions by $8\%$.

Energy and momentum scale uncertainties are estimated by assigning a 4\% uncertainty on the jet energy scale, a 1\%~(3\%) uncertainty on the electron energy scale for the barrel~(endcap) region of ECAL, and  a 1\% uncertainty on the muon momentum scale.
The effect of electron energy, muon momentum, and jet energy resolution on expected signal
and backgrounds is assessed by smearing the electron energy by 1\% and 3\% in the barrel and endcaps,
respectively, the muon momentum by 4\%, and by varying the jet energy resolution by an $\eta$-dependent value in the range 5-14\%.
In the $\ell\nu jj$ analyses, the uncertainty on the energy and momentum scales and resolutions
are propagated to the measurement of $\MET$. The effect of these uncertainties is calculated
for the (minor) background sources for which no data rescaling is applied.
For the background sources for which data rescaling is applied, residual uncertainties are calculated (\ie relative to the initial selection used to derive the rescaling factor).

Recent measurements of the muon reconstruction, identification, trigger, and
isolation efficiencies using $\cPZ\to\mu\mu$ events show very good agreement between data and MC events~\cite{zprimempas}.
A $\sim 1\%$ discrepancy is observed in the data-to-MC comparison of the muon trigger efficiency. This discrepancy is taken as a systematic uncertainty per muon, assigned to both signal and estimated background. The electron trigger and reconstruction and identification uncertainties contribute 3\% (4\%) to the uncertainty in both signal and estimated background for the $\Pe\Pe jj$ ($\Pe\nu jj$) channel.

The systematic uncertainties and their effects on signal and background
are summarized in Table~\ref{tab:SysUncertainties_all} for all channels, corresponding to the final selection optimized for $M_{\mathrm {LQ}}=600$\GeV.

\begin{table*}[htbp]
\begin{center}
\topcaption{Systematic uncertainties and their effects on signal ($S$) and background ($B$) in all channels for the $M_{LQ}=600$\GeV final selection. All uncertainties are symmetric.}
\footnotesize
\begin{tabular}{|lccccccccc|}
\hline
	 Systematic  & Magnitude  & \multicolumn{2}{c}{$\Pe\Pe jj$} & \multicolumn{2}{c}{$\mu\mu jj$}  & \multicolumn{2}{c}{$\Pe\nu jj$} & \multicolumn{2}{c|}{$\mu\nu jj$}\\
	 Uncertainties                  &  ($\%$)   &$S (\%)$    & $B (\%)$   & $S (\%)$     & $B (\%)$     & $S (\%)$   & $B (\%)$   & $S (\%)$   & $B (\%)$  \\
	 \hline
	 Jet Energy Scale               & $4$       	& $2$     	& $1$     & $1$       & $1$      & $5$    	& $8.5$ 	& $3$    	& $7$ \\
	 Background Modeling            & $-$         	& $-$       & $11$    & $-$       & $9$      & $-$    	& $11$   	& $-$       & $10$   \\
	 Electron Energy Scale          & $1(3)$    	& $1$    	& $6$     & $-$       & $-$      & $1.5$  	& $4.5$ 	& $-$       & $-$      \\
         Muon Momentum Scale   		& $1$       	& $-$       & $-$     & $0.5$     & $4$      & $-$    	& $-$      	& $1$    	& $2$   \\
         Muon Reco/ID/Iso      		& $1$       	& $-$       & $-$     & $2$       & $-$      & $-$    	& $-$      	& $1$    	& $-$      \\
	 Jet Resolution        			& $(5-14)$  	& $0.5$   	& $0.5$   & $<0.5$    & $<0.5$    & $<0.5$   & $2$    	& $<0.5$  	& $2.5$  \\
	 Electron Resolution   			& $1(3)$    	& $0.5$   	& $1$     & $-$       & $-$      & $< 0.5$  & $1.5$  	& $-$       & $-$      \\
         Muon Resolution       		& $4$       	& $-$       & $-$     & $< 0.5$   & $5$      & $-$      & $-$      	& $< 0.5$ 	& $1$    \\
         Pileup                		& $8$       	& $1$     	& $1$     & $0.5$     & $< 0.5$  & $1$      & $1.5$  	& $1$     	& $4$  \\
	 Integrated Luminosity 			& $2.2$     	& $2.2$   	& $-$     & $2.2$     & $-$      & $2.2$    & $-$       & $2.2$   	& $-$      \\ \hline
	 Total                          &             	& $3$     	& $13$    & $3$       & $11$      & $6$      & $15$   	& $4$     	& $13$   \\
\hline
\end{tabular}
\label{tab:SysUncertainties_all}
\end{center}
\end{table*}

\section{Results}
\label{results}
The number of observed events in data passing the full selection criteria is consistent with the SM background prediction in all decay channels.
An upper limit on the leptoquark pair-production cross section is therefore set using
the CL$_{\text{S}}$ modified frequentist approach~\cite{Junk:1999kv,Read:2000ru}. A log-normal probability function is used to integrate over the systematic uncertainties.
Uncertainties of statistical nature are described with $\Gamma$ distributions with
widths determined by the number of events simulated in MC samples or observed in data control regions.

The  95\%~CL upper limits on $\sigma \times \beta^2$
or $\sigma \times 2\beta(1-\beta)$ as a function of leptoquark mass are shown
together with the NLO predictions for the scalar leptoquark pair-production cross section in Figs.~\ref{fig:limit_plots_e} and~\ref{fig:limit_plots_mu}.
The theoretical cross sections are
represented for different values of the renormalization and factorization scale, $\mu$,
varied between half and twice the leptoquark mass (blue shaded region).
The PDF uncertainties are taken into account in the theoretical cross section values.

By comparing the observed upper limit with the theoretical cross section values, first-generation scalar leptoquarks with masses less than 830 (640)\GeV are excluded with the assumption that $\beta = 1~(0.5)$.
Similarly, second-generation scalar leptoquarks with masses less than 840 (620)\GeV are excluded for $\beta = 1~(0.5)$. This is to be compared with median expected limits of 790 (640)\GeV for first-generation scalar leptoquarks, and 800 (610)\GeV for second-generation scalar leptoquarks.

The observed and expected limits on the branching fraction $\beta$ as a function of leptoquark mass can be further improved using the combination of the
$\ell\ell jj$ and $\ell\nu jj$ channels, as shown in Figure~\ref{fig:limit_plots_lqcombo}. These combinations lead to the exclusion of first- and second-generation scalar leptoquarks with masses less than 640 and 650\GeV for $\beta = 0.5$, compared with median expected limits of 680 and 670\GeV.

\begin{figure}[htbp]
  \begin{center}
  \includegraphics[width=0.48\textwidth]{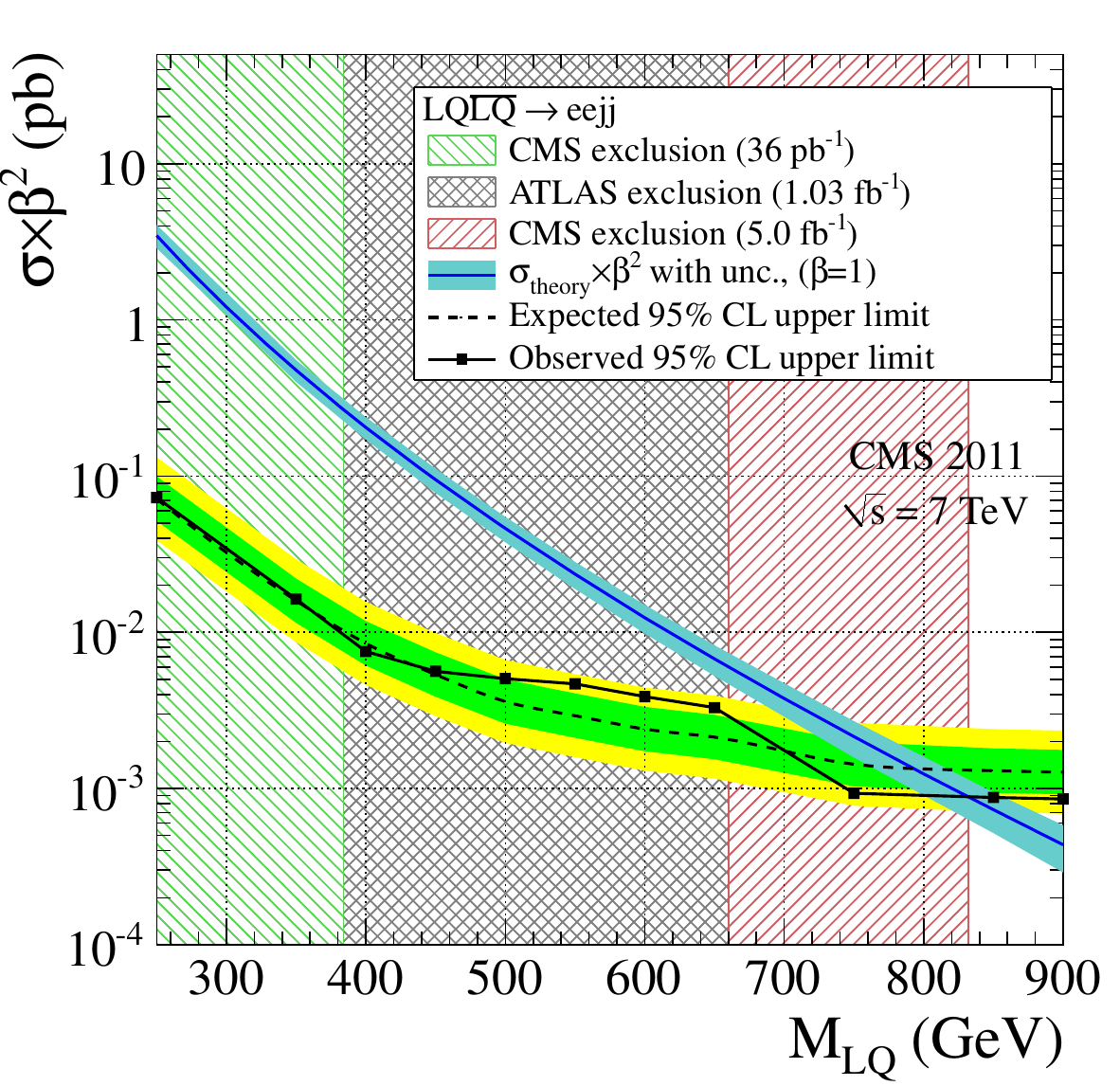}
  \includegraphics[width=0.48\textwidth]{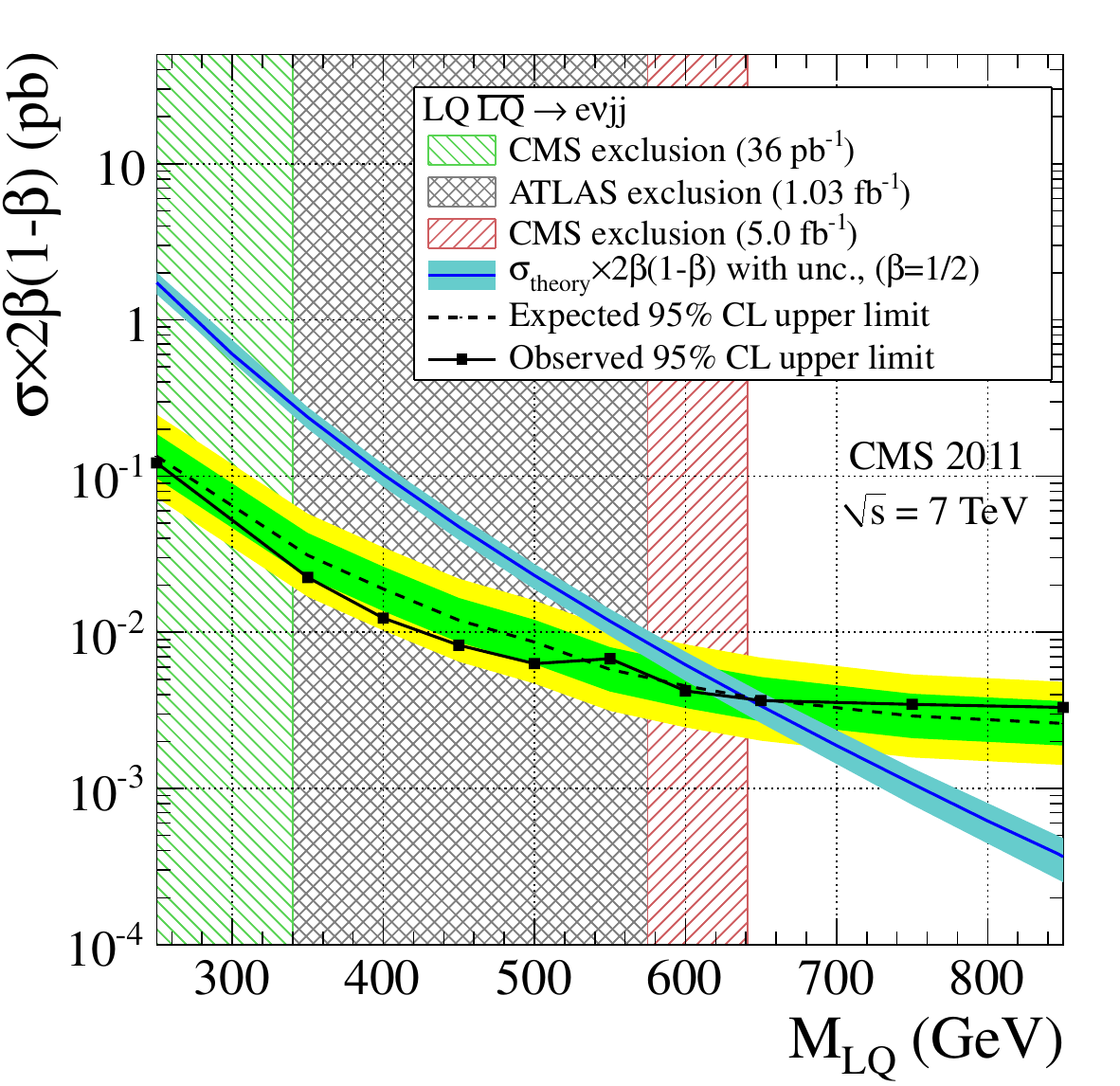}
   \caption{\cmsLLeft\ (\cmsRight) frame: the expected and observed upper limits at $95\%$ CL on the leptoquark pair-production
    cross section times $\beta$ ($2\beta(1-\beta)$) as a function of the first-generation leptoquark mass obtained with the $\Pe\Pe jj$ ($\Pe\nu jj$) analysis. The expected limits and uncertainty bands represent the median expected limits and the 68\% and 95\% confidence intervals.
    The systematic uncertainties reported in Table~\ref{tab:SysUncertainties_all} are included in the calculation.
    The left and middle shaded regions are excluded by the current ATLAS limit~\cite{ATLASLQ1} and CMS limits~\cite{CMSLQ1,CMSLQ2} for $\beta=1 (0.5)$ in the $\Pe\Pe jj$ ($\Pe\nu jj$) channel. The right shaded regions are excluded by these results.
    The $\sigma_{\rm theory}$ curves and their bands represent, respectively, the theoretical scalar leptoquark pair-production cross section
    and the uncertainties due to the choice of PDF and renormalization/factorization scales.}
  \label{fig:limit_plots_e}
  \end{center}
\end{figure}

\begin{figure}[htbp!]
  \centering
  \includegraphics[width=0.48\textwidth]{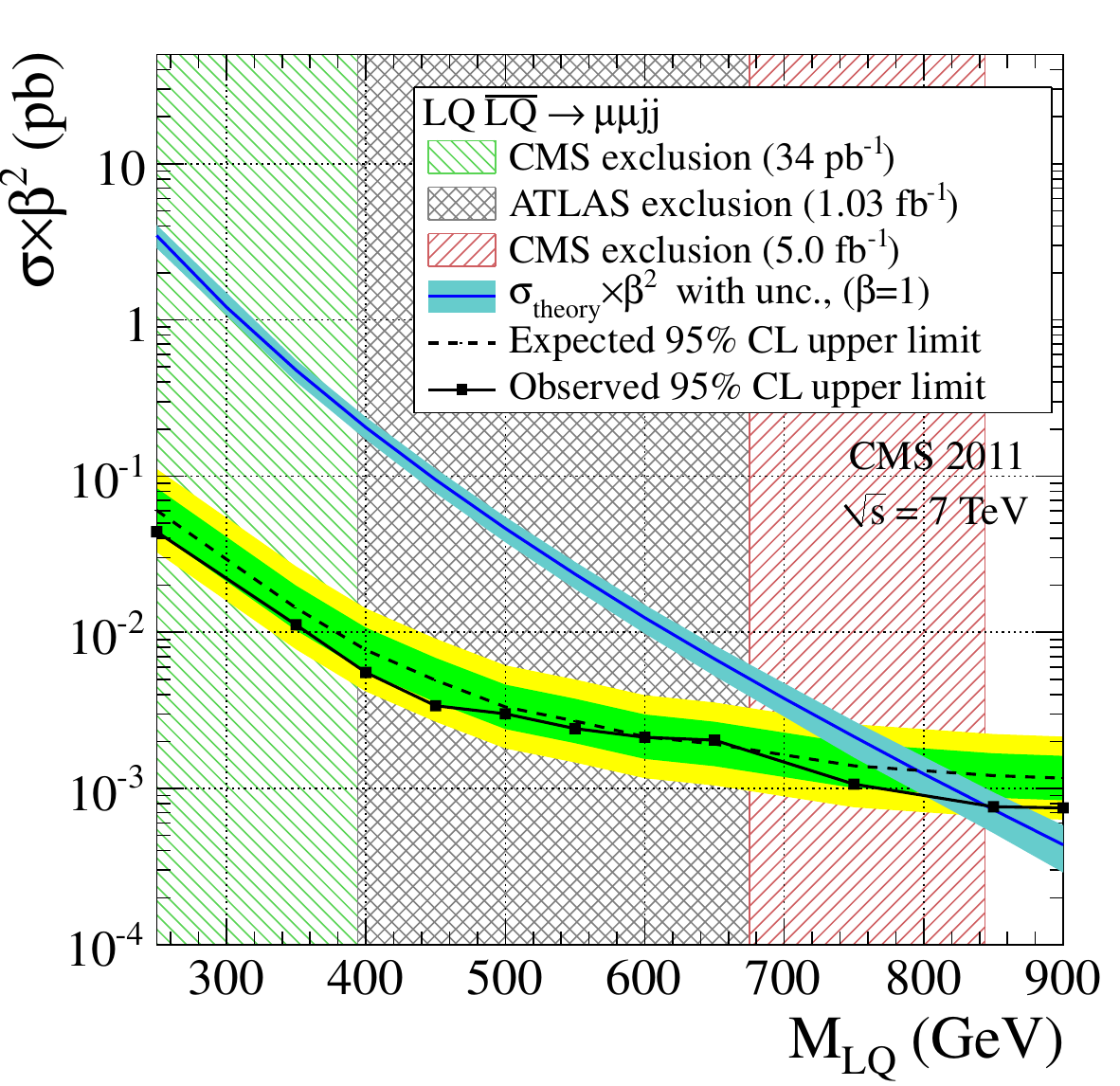}
  \includegraphics[width=0.48\textwidth]{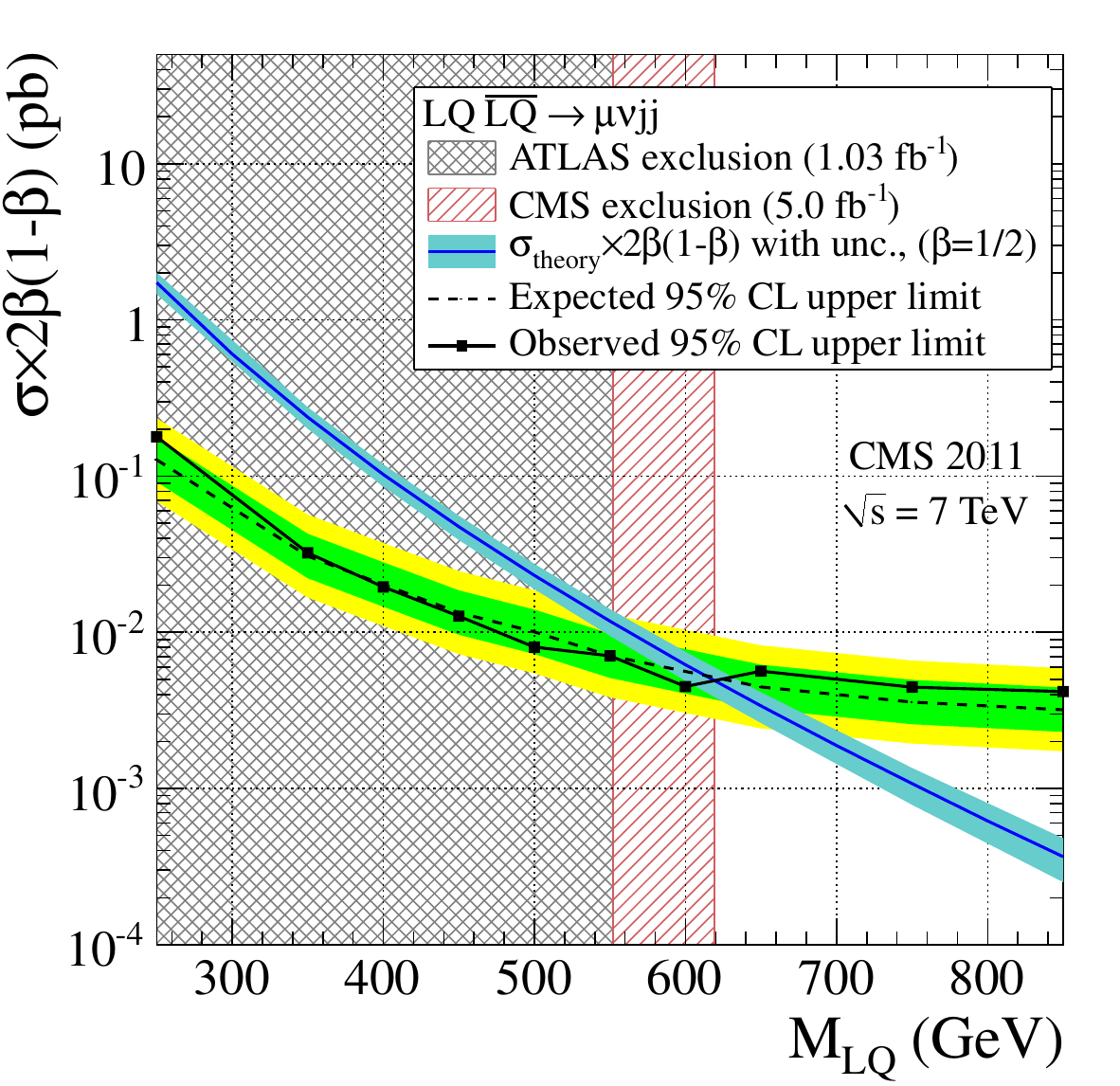}
   \caption{
    \cmsLLeft\ (\cmsRight) frame: the expected and observed upper limits at $95\%$ CL on the leptoquark pair-production
    cross section times $\beta$ ($2\beta(1-\beta)$) as a function of the second-generation leptoquark mass obtained with the $\mu\mu jj$ ($\mu\nu jj$) analysis. The expected limits and uncertainty bands represent the median expected limits and the 68\% and 95\% confidence intervals.
    The systematic uncertainties reported in Table~\ref{tab:SysUncertainties_all} are included in the calculation.
    The left and middle shaded regions in the \cmsLeft\ frame are excluded by the current ATLAS limit~\cite{ATLASLQ2} and CMS limit~\cite{CMSLQ} for $\beta=1$ in the $\mu\mu jj$ channel only. The left shaded region in the \cmsRight\ frame is excluded by the current ATLAS limit~\cite{ATLASLQ2} for $\beta=0.5$ in the $\mu\nu jj$ channel only. The right shaded regions are excluded by these results.
    The $\sigma_{\rm theory}$ curves and their bands represent, respectively, the theoretical scalar leptoquark pair-production cross section
    and the uncertainties due to the choice of PDF and renormalization/factorization scales.% \cite{PhysRevD.71.057503}.
    }
  \label{fig:limit_plots_mu}
\end{figure}

\begin{figure}[htbp!]
  \centering
  \includegraphics[width=0.48\textwidth]{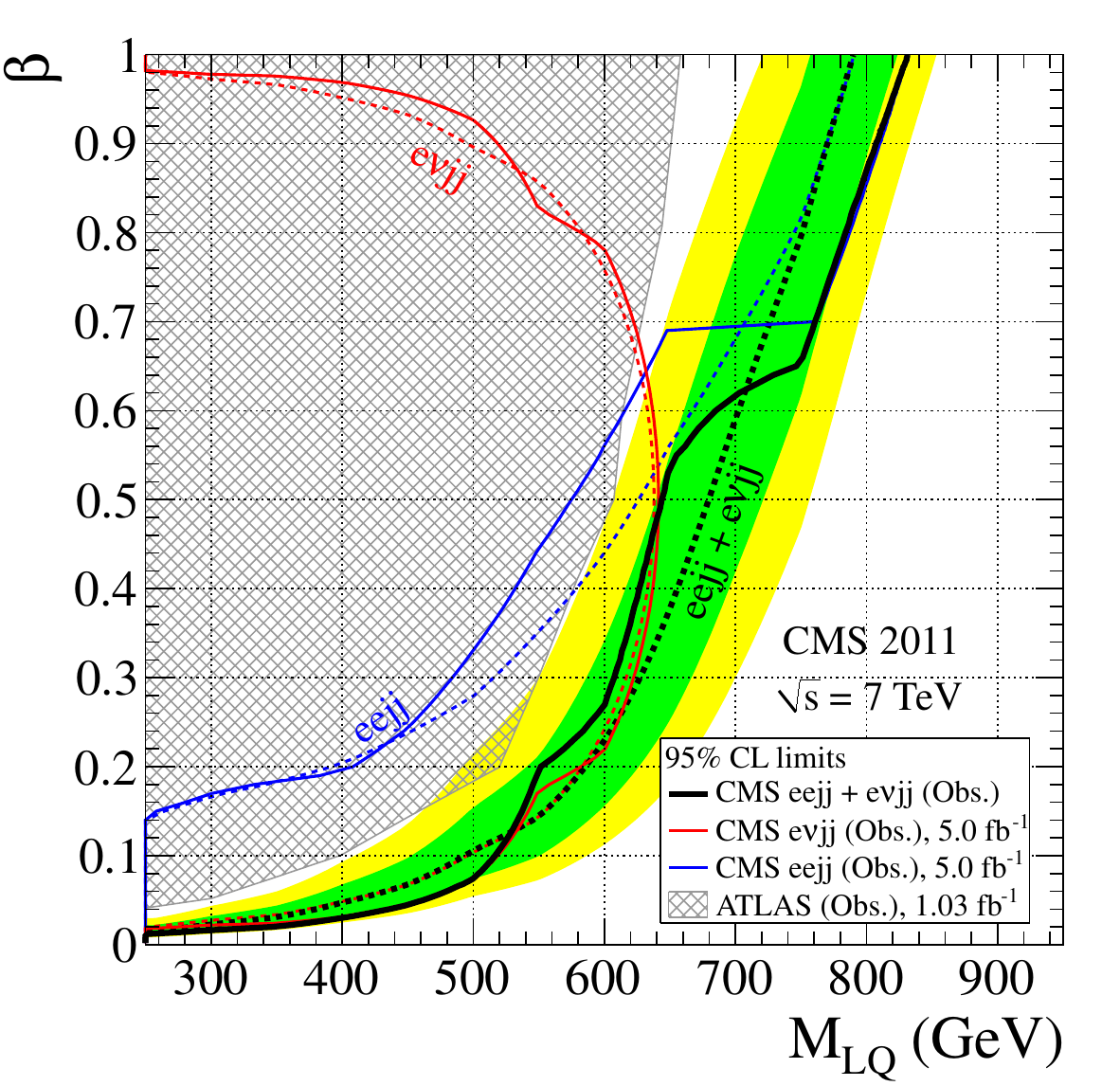}
  \includegraphics[width=0.48\textwidth]{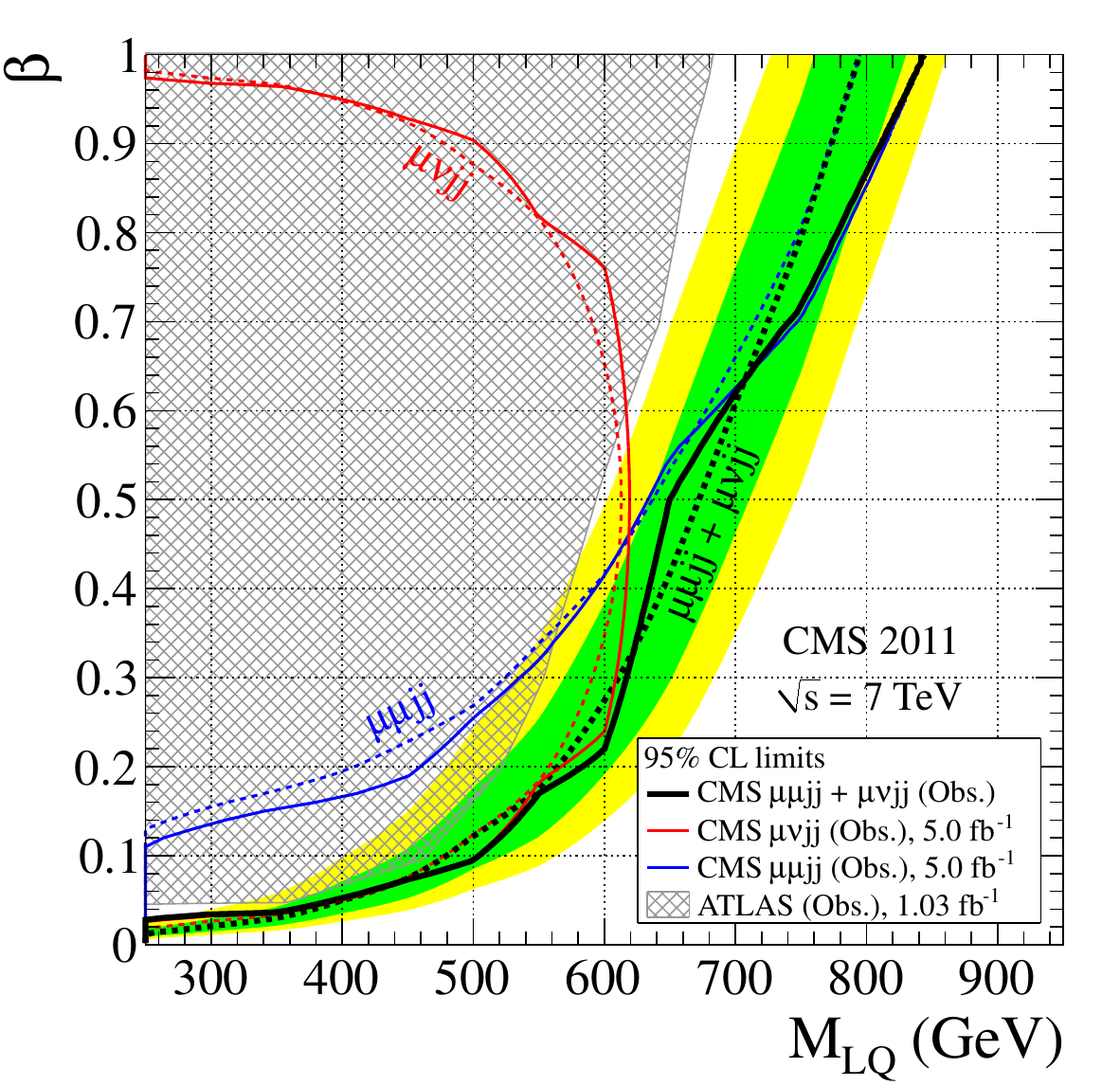}
   \caption{
    \cmsLLeft\ (\cmsRight) frame: the expected and observed exclusion limits at $95\%$ CL on the first- (second-)generation leptoquark hypothesis in the $\beta$ versus mass plane using the central value of signal cross section for the individual $\Pe\Pe jj$ and $\Pe \nu jj$ ($\mu\mu jj$ and $\mu\nu jj$) channels and their combination.
The dark green and light yellow expected limit uncertainty bands represent the 68\% and 95\% confidence intervals. Solid lines represent the observed limits in each channel, and dashed lines represent the expected limits.
    The systematic uncertainties reported in Table~\ref{tab:SysUncertainties_all} are included in the calculation.
        The shaded region is excluded by the current ATLAS limits~\cite{ATLASLQ1,ATLASLQ2}.
    }
  \label{fig:limit_plots_lqcombo}
\end{figure}

\section{Summary}
\label{conclusions}
In summary, a search for pair production of first- and second-generation scalar leptoquarks has been performed in decay channels with either two charged leptons of the same-flavor (electrons or muons) and at least two jets, or a single charged lepton (electron or muon), missing transverse energy, and at least two jets, using 7 TeV proton-proton collisions data corresponding to an integrated luminosity of 5\fbinv.
The selection criteria  have been optimized for each leptoquark signal mass hypothesis.
The number of observed candidates for each hypothesis agree with the estimated number of background events. The CL$_\mathrm{S}$ modified frequentist
approach has been used to set limits on the leptoquark cross section times the branching fraction for the decay of a leptoquark pair. At 95\%~confidence level, the pair production of first- and second-generation leptoquarks is excluded with masses below 830~(640)\GeV and 840~(650)\GeV for $\beta = 1~(0.5)$. These are the most stringent limits to date.
\ifthenelse{\boolean{cms@external}}{\clearpage}{}
\section*{Acknowledgements}
We extend our thanks to Michael Kr$\text{\"a}$mer for providing the tools for calculation of the leptoquark theoretical cross section and PDF uncertainty.

We congratulate our colleagues in the CERN accelerator departments for the excellent performance of the LHC machine. We thank the technical and administrative staff at CERN and other CMS institutes, and acknowledge support from: BMWF and FWF (Austria); FNRS and FWO (Belgium); CNPq, CAPES, FAPERJ, and FAPESP (Brazil); MES (Bulgaria); CERN; CAS, MoST, and NSFC (China); COLCIENCIAS (Colombia); MSES (Croatia); RPF (Cyprus); MoER, SF0690030s09 and ERDF (Estonia); Academy of Finland, MEC, and HIP (Finland); CEA and CNRS/IN2P3 (France); BMBF, DFG, and HGF (Germany); GSRT (Greece); OTKA and NKTH (Hungary); DAE and DST (India); IPM (Iran); SFI (Ireland); INFN (Italy); NRF and WCU (Korea); LAS (Lithuania); CINVESTAV, CONACYT, SEP, and UASLP-FAI (Mexico); MSI (New Zealand); PAEC (Pakistan); MSHE and NSC (Poland); FCT (Portugal); JINR (Armenia, Belarus, Georgia, Ukraine, Uzbekistan); MON, RosAtom, RAS and RFBR (Russia); MSTD (Serbia); SEIDI and CPAN (Spain); Swiss Funding Agencies (Switzerland); NSC (Taipei); TUBITAK and TAEK (Turkey); STFC (United Kingdom); DOE and NSF (USA).

Individuals have received support from the Marie-Curie programme and the European Research Council (European Union); the Leventis Foundation; the A. P. Sloan Foundation; the Alexander von Humboldt Foundation; the Austrian Science Fund (FWF); the Belgian Federal Science Policy Office; the Fonds pour la Formation \`a la Recherche dans l'Industrie et dans l'Agriculture (FRIA-Belgium); the Agentschap voor Innovatie door Wetenschap en Technologie (IWT-Belgium); the Council of Science and Industrial Research, India; the Compagnia di San Paolo (Torino); and the HOMING PLUS programme of Foundation for Polish Science, cofinanced from European Union, Regional Development Fund.

\bibliography{auto_generated}

\cleardoublepage \appendix\section{The CMS Collaboration \label{app:collab}}\begin{sloppypar}\hyphenpenalty=5000\widowpenalty=500\clubpenalty=5000\textbf{Yerevan Physics Institute,  Yerevan,  Armenia}\\*[0pt]
S.~Chatrchyan, V.~Khachatryan, A.M.~Sirunyan, A.~Tumasyan
\vskip\cmsinstskip
\textbf{Institut f\"{u}r Hochenergiephysik der OeAW,  Wien,  Austria}\\*[0pt]
W.~Adam, E.~Aguilo, T.~Bergauer, M.~Dragicevic, J.~Er\"{o}, C.~Fabjan\cmsAuthorMark{1}, M.~Friedl, R.~Fr\"{u}hwirth\cmsAuthorMark{1}, V.M.~Ghete, J.~Hammer, N.~H\"{o}rmann, J.~Hrubec, M.~Jeitler\cmsAuthorMark{1}, W.~Kiesenhofer, V.~Kn\"{u}nz, M.~Krammer\cmsAuthorMark{1}, I.~Kr\"{a}tschmer, D.~Liko, I.~Mikulec, M.~Pernicka$^{\textrm{\dag}}$, B.~Rahbaran, C.~Rohringer, H.~Rohringer, R.~Sch\"{o}fbeck, J.~Strauss, A.~Taurok, W.~Waltenberger, G.~Walzel, E.~Widl, C.-E.~Wulz\cmsAuthorMark{1}
% ...
\vskip\cmsinstskip
\dag:~Deceased\\
1:~~Also at Vienna University of Technology, Vienna, Austria\\
% ...

\end{sloppypar}
\end{document}